%% file: main.tex
\documentclass[letterpaper,twocolumn,10pt]{article}
\usepackage{usenix}

\usepackage[numbers]{natbib}
\setcitestyle{numbers,sort&compress=false}

\usepackage{tikz}
\usepackage{amsmath}

\usepackage{amssymb}
\usepackage[normalem]{ulem}
\usepackage[table,xcdraw]{xcolor}
\usepackage{multirow}
\usepackage{caption}

\usepackage{filecontents}

\usepackage{xurl}    
\usepackage{hyperref}

\begin{document}

\date{}

\title{\Large \bf Anchors that Don’t Lift: Understanding Supply Chain Driven Kernel Lock-In and Governance-Mediated Mitigation Strategies in SOHO Devices}

\author{
{\rm Ritwik Badola}\\
\textit{IIT Madras}\\
ritwik@cse.iitm.ac.in
\and
{\rm Rajdeep Ghosh}\\
\textit{IIT Kharagpur}\\
ghoshrajdeep2000@gmail.com
\and
{\rm Ashita Gupta}\\
\textit{IIIT Kottayam}\\
guptashita1405@gmail.com
\and 
{\rm Chester Rebeiro }\\
\textit{IIT Madras}\\
chester@cse.iitm.ac.in
\and
{\rm Mainack Mondal}\\
\textit{IIT Kharagpur}\\
mainack@cse.iitkgp.ac.in
} 

\maketitle

\begin{abstract}
\input{abstract}
\end{abstract}

\section{Introduction}
\input{introduction}

\section{Background}
\input{background}

\section{Related work}
\input{related_work}

\section{Large-scale collection of SOHO firmware and discovering supply chain}
\input{data_collection}

\section{Precise measurement of actual CVEs in SOHO firmware kernels (RQ1)}

\input{RQ/newRQ1}

\section{Tracing CVEs into the supply chain (RQ2)}

\input{RQ/newRQ2}

\section{Security choices of non-SoC vendor supply chain organizations during kernel selection (RQ3)}

\input{RQ/newRQ3}

\section{Collecting data on individual, community and regulatory governance efforts at scale}
\input{datacollection_RQ4}

\section{Limitations}
\input{limitations}

\section{Effectiveness of potential mediation by individual, community and regulatory governance on kernel lock-in (RQ4)}

\input{RQ/RQ4}

\section{Implications}

\input{implications}

\section{Recommendations for Stakeholders}
\input{recommendation}

\section{Conclusion}
\input{conclusion}

\section{Acknowledgments}

We acknowledge the use of generative AI tools (ChatGPT, Gemini) for minor editorial assistance, including grammar, clarity, and stylistic refinement of the manuscript; however all technical content, analyses, and conclusions remain the responsibility of the authors.

\section{Ethical Considerations}
\input{ethical_consideration}

\section{Open Science}
\input{openscience}

\bibliographystyle{plain}
\bibliography{references}

\appendix

\section*{Appendix}

\section{Large-scale collection and analysis of SOHO kernels}

\subsection{Firmware unpacking and binary-based version-string extraction}\label{sec:binary_analysis_baseline}

At the firmware analysis stage, we attempt extraction and unpacking using EMBA~\cite{embaGitHub} as an automated first-pass for large-scale processing. EMBA is an automated firmware analysis pipeline that we use here to (i) unpack firmware images into recoverable filesystem artifacts, (ii) extract kernel image and version-strings needed to recover the shipped kernel version, and (iii) derive a version-centric upper-bound CVE candidate count for that inferred kernel line (as a “maximum possible” reference point). We treat the unpacking step as successful for our use case only when we can recover (i) a usable filesystem extraction and (ii) kernel-related artefacts sufficient to extract a kernel baseline. For kernel baseline extraction, we rely on EMBA’s S24 (\emph{Find kernel}) and S25 (\emph{Check kernel}) modules, which identify and report embedded Linux kernel version strings using linux-exploit-suggester~\cite{thez-labs_linux-exploit-suggester}. We treat this firmware-derived kernel baseline as a reference point for prior work's version-centric estimates, but we do not rely on it for the \emph{actual} inference of vulnerability presence (which we instead perform using vendor-released firmware source trees). Although EMBA is our primary unpacking pipeline for firmware images, we additionally use binwalk~\cite{binwalkGitHub} and unblob~\cite{unblobOrg} in two situations: (i) manual verification of edge cases where EMBA extraction results appear incomplete or inconsistent, and (ii) a small Commercial Off-The-Shelf (COTS) validation subset, where firmware is extracted directly from devices and benefits from additional unpacking confirmation across tools. This binary-based side analysis provides a baseline comparable to existing large-scale approaches and serves as the firmware reference point for later correlation with firmware source releases.

\subsection{Extraction and robust matching of SDK-derived kernel baseline and firmware-derived kernel baseline}\label{app_sec4}

We perform consistency checks to support the claim that upstream platform baselines are reflected in shipped firmware, and to bound mismatches between firmware and GPL artefacts. For each device where we can infer an SDK kernel baseline from the GPL tree, we extract the Linux kernel baseline from the corresponding firmware binary and compare the two at a normalised baseline granularity (X.Y.Z). An agreement is treated as evidence that the shipped firmware follows the same baseline kernel track as the GPL release, providing a concrete link between GPL-derived platform lineage and what is deployed in firmware.

To reduce the possibility that observed baseline alignment is coincidental or device-specific, we additionally examine cross-vendor cases where the same SoC appears in devices from different vendors. For these cases, we compare SDK markers and GPL artefacts across vendors and cross-check that firmware-extracted kernel baselines remain consistent across devices sharing the same SoC. We analyse 77 such cross-vendor device cases and observe that devices from different vendors using the same SoC often rely on the same Linux kernel baseline, consistent with shared platform dependencies rather than isolated per-device kernel selection. NDA-constrained platforms remain a recurring challenge because public SoC/SDK information is limited or absent in these cases, making GPL artefacts the primary best-effort source for platform lineage and build context. We treat missing public attribution, missing GPL releases, and insufficient GPL markers as explicit coverage limits rather than silent failures. 

Finally, we include a small Commercial Off-The-Shelf (COTS) validation subset consisting of an Amazon's Choice SOHO router and an IP camera. For each device, we extract firmware directly from the SOP8 flash package using a CH341A programmer, dump the flash contents to a host machine, and run the resulting firmware image through the same extraction and correlation pipeline as the dataset-derived firmware. We then retrieve the corresponding GPL releases from vendor portals and verify that the same baseline inference and correlation steps apply, providing a sanity check that our dataset-driven observations remain consistent with real retail devices. We keep COTS device identities anonymous and use this subset only for validation rather than getting corpus-wide conclusions.

\section{Data Collection to analyse Governance efforts around SOHO devices }\label{app_rq4_dc}

\subsection{Collecting data from Reddit}\label{app_rq4_dc_userperception}

Reddit users refer to devices using a variety of lexical forms, including differences in spacing, hyphenation, and spelling. To account for this variability, for each device we constructed multiple search queries consisting of the canonical device name and additional variants derived from common naming alternatives observed in practice. For each query variant, we issued both lowercase and uppercase versions to further improve coverage. For example, for \textit{D-LINK DIR-825}, we issued queries using \textit{D-LINK DIR-825}, \textit{DLINK DIR-825}, and \textit{D-LINK DIR825}, along with their corresponding lowercase forms. Reddit's public search interface allows for post search, but does not offer an API for large-scale data collecting. Hence, we used browser automation via Selenium~\cite{gundecha2018selenium} to emulate user interactions with Reddit’s search functionality. We queried Reddit for each device, collecting 4,104 unique posts across 306 devices and their query variants resulting in 22,875 unique comments across 4,104 Reddit posts.

\subsection{Leveraging ML to identify security and privacy related comments}\label{app_rq4_dc_ml}

\noindent
\textbf{Creating ground truth labels: } In particular, our goal is to distinguish security and privacy related comments from comments not related to security and privacy among the large set of collected Reddit comments. To that end, we manually created a gold-standard dataset consisting of 762 comments with two classes: related to security and privacy and not related to security and privacy as follows. First, we randomly sampled around 1000 comments from the keyword-filtered subset of comments that potentially indicate security or privacy-related concerns. Furthermore, we randomly sampled 500 more comments from the remaining comments that did not match any of the security or privacy related keywords. This sampling strategy ensured that the labeled dataset contained a balanced representation of both candidate security and privacy related and clearly non security and privacy related comments. Two coders independently coded these 1500 comments using two labels: related to security and privacy and not related to security and privacy. The annotators achieved a Cohen’s kappa of 0.79, indicating substantial agreement. The annotators then met to resolve disagreements and assign final labels. To handle comments that contained multiple sentences or discussed multiple topics, we applied a conservative rule: if at least one sentence or coherent thought within a comment expressed a security- or privacy-related concern, we labeled the entire comment as security and privacy related. Total we had 362 security and privacy related comments and 1138 non-security-related. We took 400 from 1138 and constituted the final labeled dataset consisting of 762 comments(362 security and privacy related comments and 400 non security and privacy related comments). We model the task of identifying security and privacy related Reddit comments as a two-class text classification problem, where the comment text serves as input to the classifier. Using the manually labeled dataset as ground truth, we experimented with multiple fine-tuned transformer-based BERT models. Among the evaluated models, we selected a fine-tuned DeBERTa for further analysis, as it demonstrated the best overall performance. In particular, the selected model achieved an F1-score of 0.95 (Detailed performance metrics for the evaluated models are reported in Table~\ref{comparative_model_performance})

\vspace{1mm}
\noindent
\textbf{Model Architecture and Training: }Given the nature of our task and the availability of manually labeled data, we utilized a range of pretrained transformer-based models to address the classification challenge. Specifically, we evaluated BERT~\cite{devlin2018bert}, RoBERTa~\cite{liu2019roberta} and DeBERTa~\cite{he2020deberta}, FLAN-T5~\cite{chung2024scaling} all of which were fine-tuned on our labeled dataset for the downstream task of comment classification (details of model architecture given in Table~\ref{details_bertmodels_arch} 
. We leveraged systematic hyperparameter tuning using the Optuna framework~\cite{akiba2019optunanextgenerationhyperparameteroptimization} to identify optimal configurations for each model. We achieved the best accuracy for the downstream task using the fine-tuned DeBERTa model---class-wise performance of the DeBERTa model is shown in Table~\ref{best-model-classwise-performance} 
. Since the task of identifying security and privacy related comments is a sequence classification task and hence, we used different architectures of models described in Table~\ref{details_bertmodels_arch} 
along with the specification of model sizes.

\begin{table}[]
\scriptsize
\centering
\caption{ Confusion matrix for our DeBERTa text-classifier annotation and ground truth for previously unseen 100 comments. For only 6, comments related to security and privacy is misclassified as
not related to security and privacy or vice-versa (marked in gray)}
\begin{tabular}{rr|rr|}
\cline{3-4}
 &  & \multicolumn{2}{c|}{\textbf{Predicted}} \\ \cline{3-4} 
 & \multirow{-2}{*}{} & \multicolumn{1}{l|}{\cellcolor[HTML]{FFFFFF}\begin{tabular}[r]{@{}r@{}}Reltaed to \\ Security/Privacy\end{tabular}} & \cellcolor[HTML]{FFFFFF}\begin{tabular}[r]{@{}r@{}}Not Reltaed to \\ Security/Privacy\end{tabular} \\ \hline
\multicolumn{1}{|r|}{} & \cellcolor[HTML]{FFFFFF}\begin{tabular}[c]{@{}r@{}}Reltaed to \\ Security/Privacy\end{tabular} & \multicolumn{1}{r|}{6} & \cellcolor[HTML]{C0C0C0}5 \\ \cline{2-4} 
\multicolumn{1}{|r|}{\multirow{-2}{*}{\textbf{\begin{tabular}[r]{@{}r@{}}Ground \\ Truth\end{tabular}}}} & \cellcolor[HTML]{FFFFFF}\begin{tabular}[r]{@{}r@{}}Not Reltaed to \\ Security/Privacy\end{tabular} & \multicolumn{1}{r|}{\cellcolor[HTML]{C0C0C0}1} & 88 \\ \hline
\end{tabular}

\label{Miscallsification_deberta}
\end{table}

\begin{table}[]
\scriptsize
\centering
\caption{: Class-wise performance of our fine-tuned DeBERTa
model on the validation dataset.}
\begin{tabular}{|r|r|r|r|}
\hline
\rowcolor[HTML]{C0C0C0} 
 & \textbf{Precison} & \textbf{Recall} & \textbf{F1 score} \\ \hline
\cellcolor[HTML]{C0C0C0}\textbf{\begin{tabular}[r]{@{}r@{}}Comments \\related to Security/Privacy\end{tabular}} & 0.92 & 0.97 & 0.94 \\ \hline
\cellcolor[HTML]{C0C0C0}\textbf{\begin{tabular}[r]{@{}r@{}}Comments not \\related to Security/Privacy\end{tabular}} & 0.97 & 0.93 & 0.95 \\ \hline
\end{tabular}

\label{best-model-classwise-performance}
\end{table}

\noindent \textbf{Misclassification analysis}: To further investigate the performance and misclassification of the DeBETa model, we randomly sampled an additional 100 Reddit comments that were not part of the training or validation sets. One annotator manually labeled these comments to create a ground-truth set. We then applied our trained classifier to these comments and constructed a confusion matrix (see Table~\ref{Miscallsification_deberta}) 
. We note that, in this confusion matrix, only 6 comments were misclassified between comments related to security and privacy and comments not related to security and privacy---five comments related to security and privacy as comments not related to security and privacy, and only one comment not related to security and privacy as security and privacy related. Thus, we identify comments related to security and privacy correctly in 95\% of cases.

\subsection{Identification of security and privacy related themes}\label{app_rq4_dc_themes}

\vspace{1mm}
\noindent
\textbf{Affinity diagramming:} Following open coding, the two researchers collaboratively analyzed the resulting codes using affinity diagramming. They did this by looking at the collection of quotes for each code in addition to the code itself~\cite{10.1145/2702123.2702561}. To assess saturation of the thematic hierarchy, we initially set aside 10 randomly selected codes. The remaining codes were then iteratively grouped into higher-level themes. This process was repeated over multiple rounds until the researchers agreed that no new higher-level themes were emerging. At the end of this process, we identified a 4-level hierarchy of themes, where Level-1 contains abstract highest-level themes created in the final round of affinity diagramming, and Level-4 contains the very specific codes specified by the open coding step. Finally, we verified that incorporating the 10 random codes did not add any new Level-1 and Level-2 themes, indicating thematic saturation of affinity diagramming.

\vspace{1mm}
\noindent
\textbf{Details about the theme hierarchy:} The entire theme hierarchy is given in Table~\ref{four_level_hierarchy} 
At Level~1, the hierarchy consists of a single overarching theme capturing security and privacy aspects of SOHO environments. Level~2 refines this theme into several high-level categories, including configuration issues, network security, software-related concerns, awareness-related discussions, update cycles, WiFi protocols, vulnerabilities of other IoT devices, and miscellaneous topics. Level~3 further decomposes these categories into more specific concerns such as data sharing, multi-device setups, firewall configurations, credential management, missing patches, DNS-related issues, and device interoperability. We note that our qualitative analysis so far is based on a subset of manually analyzed Reddit comments identified as security and privacy related.  For additional clarity, Table~\ref{description-all-reddit-comments} 
details the Level-2 categories by grouping comments based on their relevance to SOHO security and privacy concerns and providing brief descriptions of each theme.

\section{Supplementary Analysis of Developer/Vendor Forums}\label{other_than_reddit}

Our primary analysis (Section-~\ref{sec_8_1}) uses Reddit to capture general user perception of SOHO device security and privacy. Since technical discussions around firmware and kernel-related issues may be more prevalent in dedicated forums, we extended our analysis to assess whether our thematic classification generalizes and whether new themes emerge from the posts and comments from these developer forums.

\begin{table}[]
\scriptsize
\centering
\caption{ Distribution of security and privacy related posts across developer and vendor forums}

\begin{tabular}{|
>{\columncolor[HTML]{C0C0C0}}c |c|c|}
\hline
\textbf{Forums} & \cellcolor[HTML]{C0C0C0}\textbf{\#total comments} & \cellcolor[HTML]{C0C0C0}\textbf{\begin{tabular}[c]{@{}c@{}}\#comments related \\ to security/privacy\end{tabular}} \\ \hline
DD-WRT          & 10000                                             & 1058                                                                                                               \\ \hline
OpenWrt         & 10000                                             & 1408                                                                                                               \\ \hline
SNB             & 5118                                              & 545                                                                                                                \\ \hline
Netgear         & 743                                               & 75                                                                                                                 \\ \hline
TP-Link         & 76                                                & 3                                                                                                                  \\ \hline
\textbf{Total}  & \textbf{25937}                                             & \textbf{3089}                                                                                                \\ \hline
\end{tabular}
\label{table_otherreeddit}
\end{table}

\vspace{1mm}
\noindent
\textbf{Forum Data Collection.} To complement our Reddit analysis, we collected posts from five dedicated forums: OpenWrt, DD-WRT, SNB, Netgear, and TP-Link developer forums. We collected up to 10,000 of the most recent posts per forum, yielding a total of 25,937 posts across all five platforms(Details given in Table-~\ref{table_otherreeddit}).

\vspace{1mm}
\noindent
\textbf{Classifier Performance Across Forums.} We applied our fine-tuned DeBERTa classifier (Section-~\ref{sec_8_1_1}) to the collected forum posts. Security and privacy related discussions account for 11.8\% of posts in this combined dataset, compared to ~5.4\% on Reddit; a modest but expected increase given the more technically oriented nature of these communities.

\vspace{1mm}
\noindent
\textbf{Thematic Findings.} To assess thematic coverage, two independent annotators manually coded 100 randomly sampled posts per forum using the codebook developed in Section-~\ref{sec_8_1_2}, achieving an inter-rater agreement (Cohen's kappa) of 0.768. The annotation reveals two key findings: (1) all discussions map to existing Level-2 themes with no new themes emerging, confirming thematic saturation and generalizability of our coding scheme; (2) configuration issues dominate security and privacy related posts, suggesting advanced users remain focused on device-level concerns rather than broader supply-chain or kernel-level issues.

These results reinforce rather than alter our thematic classification. The consistency of themes across general and advanced-user communities confirms that the governance gap identified in Section-~\ref{sec_8_1} persists even among technically aligned users; kernel lock-in and supply-chain vulnerabilities remain outside the purview of typical user discourse regardless of the platform.

\begin{table*}[]
\scriptsize
\centering
\caption{Our four-level hierarchical themes explaining security and privacy-related discussions among Reddit users about consumer devices.}
\begin{tabular}{|c|c|c|c|}
\hline
\rowcolor[HTML]{C0C0C0} 
\cellcolor[HTML]{CBCEFB}\textbf{A. S/p of SOHO}  & A.3 Awareness                       & \cellcolor[HTML]{FFFFFF}A.6.2 MAC address spoofing                                                    & B.3 Software related                            \\ \hline
\rowcolor[HTML]{FFFFFF} 
\cellcolor[HTML]{C0C0C0}A.1 Configuration issues & A.3.1 Working setup                 & A.6.3 Privacy features                                                                                & B.3.1 DNS                                       \\ \hline
\rowcolor[HTML]{FFFFFF} 
A.1.1 Multi-device setup                         & A.3.2 Credential                    & A.6.4 Virtualization of SOHO                                                                           & B.3.2 Comparison                                \\ \hline
\rowcolor[HTML]{FFFFFF} 
A.1.2 Incorrect settings                         & A.3.3 Alternate firmware            & A.6.5 Bashing the device-vendor                                                                       & B.3.3 Host computer                             \\ \hline
\rowcolor[HTML]{FFFFFF} 
A.1.3 Device feature                             & A.3.4 Recommendation                 & A.6.6 Others                                                                                          & B.3.4 Other device firmware                     \\ \hline
\cellcolor[HTML]{D9EAD3}A.1.3.1 DNS              & \cellcolor[HTML]{C0C0C0}A.4 Exploit & \cellcolor[HTML]{CBCEFB}\textbf{B. Not S/p of SOHO}                                                   & \cellcolor[HTML]{FFFFFF}B.3.5 Remote management \\ \hline
\rowcolor[HTML]{FFFFFF} 
\cellcolor[HTML]{D9EAD3}A.1.3.2 VPN              & A.4.1 Via credentials               & \cellcolor[HTML]{C0C0C0}B.1 WiFi Protocol                                                             & B.3.6 Vendor-app                                \\ \hline
\rowcolor[HTML]{FFFFFF} 
\cellcolor[HTML]{D9EAD3}A.1.3.3 IPV-6            & A.4.2 Known attack                  & B.1.1 Impact                                                                                          & B.3.6 Proxy                                     \\ \hline
\rowcolor[HTML]{FFFFFF} 
\cellcolor[HTML]{D9EAD3}A.1.3.4 MAC              & A.4.3 Vulnerability disclosure      & B.1.2 Compatibility                                                                                   & B.3.7 Others                                    \\ \hline
\rowcolor[HTML]{C0C0C0} 
A.2 Network security                             & A.5 Update cycle                    & B.2 Configuration                                                                                     & B4 Vulnerability of other iot                   \\ \hline
\rowcolor[HTML]{FFFFFF} 
A.2.1 Firewall                                   & A.5.1 Missing patches               & B.2.1 Alternate resource                                                                              & \cellcolor[HTML]{C0C0C0}B.5 Misc                \\ \hline
\rowcolor[HTML]{FFFFFF} 
A.2.2 Network monitoring                         & A.5.2 Frequency                     & B.2.2 Issues related to configuration                                                                 & B.5.1 Company Policy                            \\ \hline
\rowcolor[HTML]{FFFFFF} 
\cellcolor[HTML]{D9EAD3}A.2.1 Parental control   & A.5.3 Company policy                & \begin{tabular}[c]{@{}c@{}}B.2.3 Recommendation/suggestion \\ to alternate configuration\end{tabular} & B.5.2 Others                                    \\ \hline
\rowcolor[HTML]{FFFFFF} 
\cellcolor[HTML]{D9EAD3}A.2.2 Techniques         & \cellcolor[HTML]{C0C0C0}A.6 Misc    & B.2.4 Help guide to configure                                                                         &                                                 \\ \hline
\rowcolor[HTML]{FFFFFF} 
\cellcolor[HTML]{D9EAD3}A.2.3 Prevention         & A.6.1 Data sharing concerns         & B.2.5 Awareness about configuration                                                                   &                                                 \\ \hline
\end{tabular}

\label{four_level_hierarchy}
\end{table*}

\begin{table*}[]
\scriptsize
\centering
\caption{Specification and architecture of models used in the classification of Reddit comments}
\begin{tabular}{|c|c|c|c|}
\hline
\rowcolor[HTML]{C0C0C0} 
{\color[HTML]{000000} \textbf{Model}}    & {\color[HTML]{000000} \textbf{Model Specification}} & {\color[HTML]{000000} \textbf{Classification of issues}} & \textbf{Theme Identification}                                                                          \\ \hline

{\color[HTML]{000000} \textbf{BERT}}     & {\color[HTML]{000000} BERT-base (Uncased)}          & {\color[HTML]{000000} Sequence classification}           & \begin{tabular}[c]{@{}c@{}}Sequence Classification\\ (With multiple classification heads)\end{tabular} \\ \hline
\textbf{RoBERTa}                         & RoBERTa-base                                        & Sequence classification                                  & \begin{tabular}[c]{@{}c@{}}Sequence Classification\\ (With multiple classification heads)\end{tabular} \\ \hline
\textbf{Flan-T5}                         & Flan-T5-base                                        & Seq2Seq                                                  & Seq2Seq                                                                                                \\ \hline
\textbf{DeBERTa}                         & DeBERTa-v3-base                                     & Sequence classification                                  & \begin{tabular}[c]{@{}c@{}}Sequence Classification\\ (With multiple classification heads)\end{tabular} \\ \hline
\end{tabular}

\label{details_bertmodels_arch}
\end{table*}

\begin{table*}[]
\caption{Level 1 and Level-2 thematic categories derived from manual coding of Reddit discussions, along with a brief description summarizing the underlying theme.}
\begin{tabular}{|l|l|l|}
\hline
\rowcolor[HTML]{C0C0C0} 
\textbf{L1 level} & \textbf{L2 level} & \textbf{Description} \\ \hline
\rowcolor[HTML]{FFFFFF} 
\cellcolor[HTML]{FFFFFF} & Configuration issues &  \begin{tabular}[l]{@{}l@{}}Problems arising from incorrect, incomplete, or confusing device \\ configurations that might effect security and privacy of SOHO devices.\end{tabular}\\ \cline{2-3} 
\rowcolor[HTML]{FFFFFF} 
\cellcolor[HTML]{FFFFFF} & Network security &  \begin{tabular}[l]{@{}l@{}}Concerns related to protecting network traffic,  parental control,\\ and perimeter defenses.\end{tabular} \\ \cline{2-3} 
\rowcolor[HTML]{FFFFFF} 
\cellcolor[HTML]{FFFFFF} & Awareness &  \begin{tabular}[l]{@{}l@{}} Gaps in user knowledge or understanding of SOHO device \\ security and privacy practices.\end{tabular} \\ \cline{2-3} 
\rowcolor[HTML]{FFFFFF} 
\cellcolor[HTML]{FFFFFF} & Exploit &  \begin{tabular}[l]{@{}l@{}} References to active or potential exploitation of vulnerabilities \\in SOHO devices.\end{tabular} \\ \cline{2-3} 
\rowcolor[HTML]{FFFFFF} 
\cellcolor[HTML]{FFFFFF} & Update cycle &  \begin{tabular}[l]{@{}l@{}} Issues related to missing, delayed, or infrequent security patches and the \\ lack of firmware updates.\end{tabular} \\ \cline{2-3} 
\rowcolor[HTML]{FFFFFF} 
\multirow{-6}{*}{\cellcolor[HTML]{FFFFFF}\begin{tabular}[c]{@{}l@{}}Related to S/P\\ of SOHO devices\end{tabular}} & Misc & \begin{tabular}[l]{@{}l@{}} Discussions that touch on SOHO security or privacy but do not \\ fit into a specific technical category.\end{tabular}  \\ \hline
\rowcolor[HTML]{FFFFFF} 
\cellcolor[HTML]{FFFFFF} & Wifi protocol &  \begin{tabular}[l]{@{}l@{}} Discussions focused on preferred wireless standards, performance, or \\ inter-operatability  rather than security or privacy.\end{tabular}\\ \cline{2-3} 
\rowcolor[HTML]{FFFFFF} 
\cellcolor[HTML]{FFFFFF} & Configuration &   \begin{tabular}[l]{@{}l@{}} General setup and configuration issues not directly tied to security \\ and privacy risks of SOHO devices\end{tabular}\\ \cline{2-3} 
\rowcolor[HTML]{FFFFFF} 
\cellcolor[HTML]{FFFFFF} & Software related &  \begin{tabular}[l]{@{}l@{}} Problems stemming from applications, host computers, or services \\ external to SOHO device security.\end{tabular} \\ \cline{2-3} 
\rowcolor[HTML]{FFFFFF} 
\cellcolor[HTML]{FFFFFF} & Vulnerability of other iot &   \begin{tabular}[l]{@{}l@{}} Security issues concerning non-SOHO IoT devices or in general about IoT\\ecosystems. \end{tabular}\\ \cline{2-3} 
\rowcolor[HTML]{FFFFFF} 
\multirow{-5}{*}{\cellcolor[HTML]{FFFFFF}\begin{tabular}[c]{@{}l@{}}Not Related to S/P\\ of SOHO devices\end{tabular}} & Misc &   \begin{tabular}[l]{@{}l@{}} Content that is not specific to any SOHO device's security and \\ privacy concerns.\end{tabular}\\ \hline
\end{tabular}

\label{description-all-reddit-comments}
\end{table*}

\section{Kernel EoL}\label{app_kernel_eol}

\begin{figure*}
\centering
    \includegraphics[width=\textwidth]{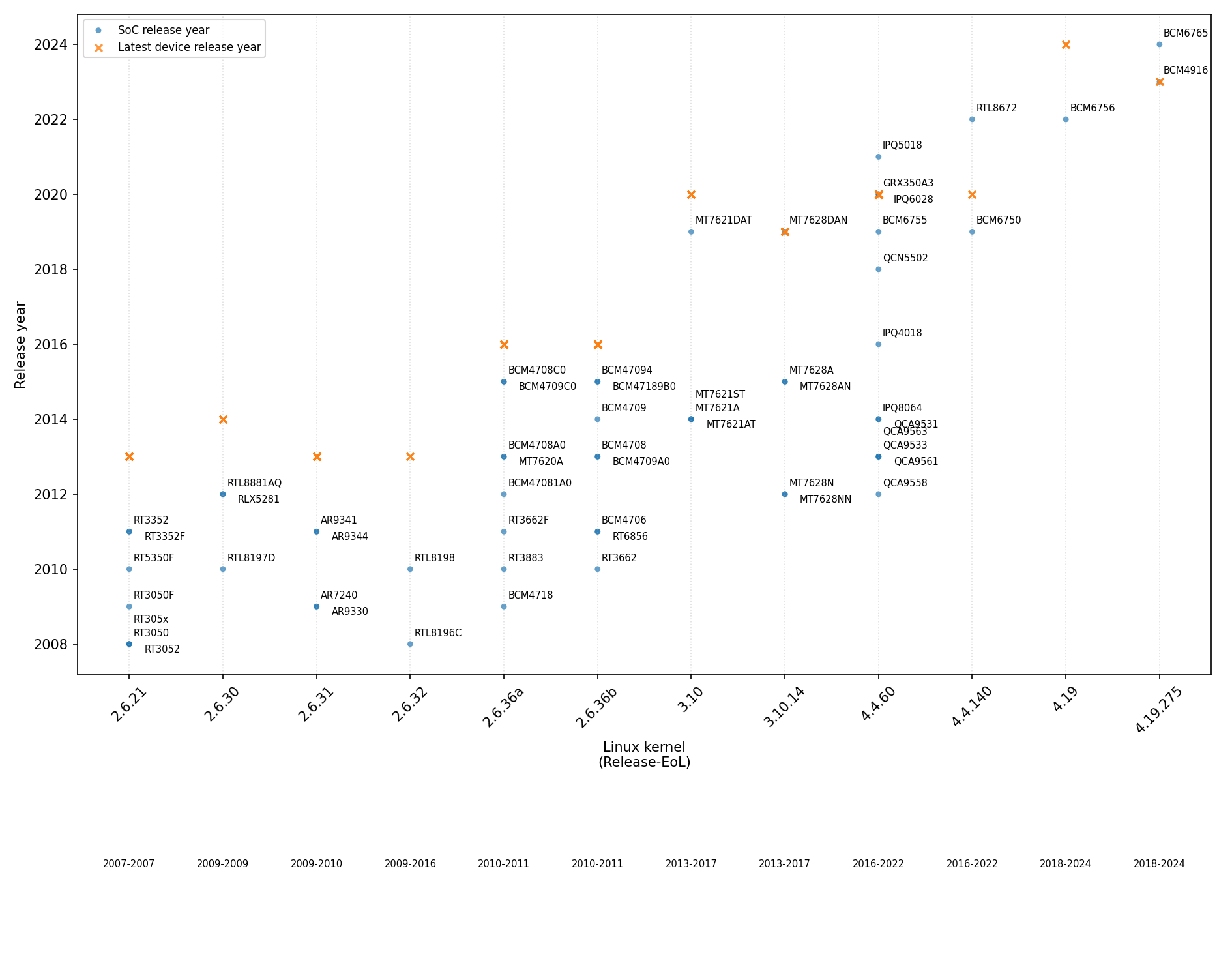}
    \caption{Graph representing how SoC and their corresponding devices use kernel baselines several years after they have reached their end of life}
\label{fig:sdk-kernel-map}
\end{figure*}

This graph (Figure-~\ref{fig:sdk-kernel-map}) represents the Linux kernel that SoC used in their SDK, the release years for those SoCs, and the release year for a device that ended up using a SoC that uses such that kernel.

\section{GPL vs Firmware Kernel}\label{app:firmware-gpl-mismatch}
We found cases where the GPL source contained a newer kernel than the firmware binary. Manual re-checks found no extraction failures; we interpret these as release mismatches between two separately maintained public artifacts — a pattern documented in prior work. This can occur because GPL sources and firmware binaries follow different publication workflows, or because devices receive updates via OTA channels not reflected in manual-download portals. We conservatively use the higher of the two observed versions, giving the device the benefit of the doubt and avoiding overstating kernel lock-in.

\section{Overall Pipeline}\label{app-large-fig}
\begin{figure*}
\centering
    \includegraphics[width=\textwidth]{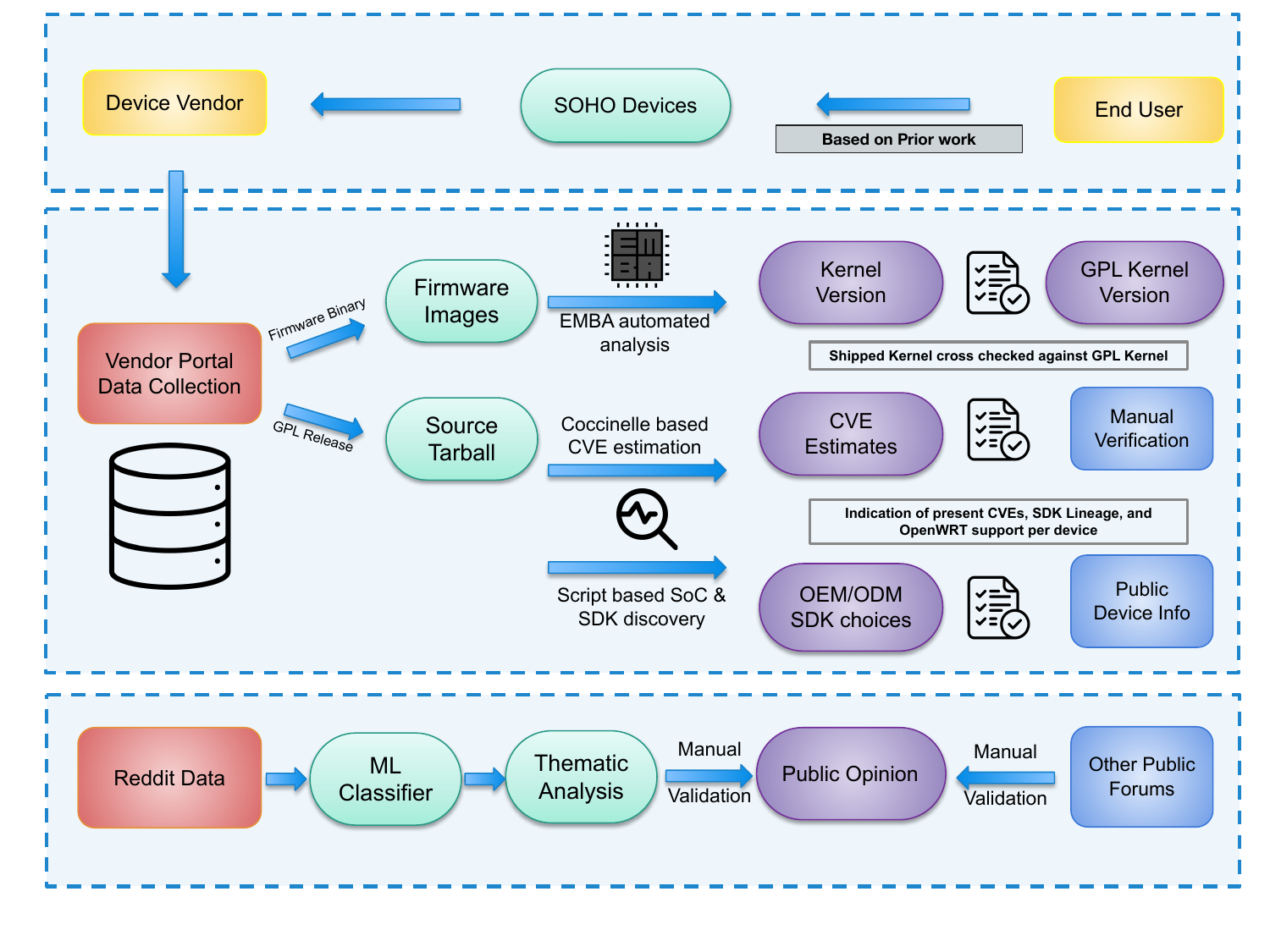}
    \caption{End-to-end SOHO Supply Chain analysis pipeline, combining automated firmware/source analysis with community opinion}
    
\label{fig:soho-pipeline-full}
\end{figure*}

Our pipeline combines automation, heuristic inference, and targeted manual validation. Automated steps include GPL/firmware collection, unpacking, kernel version extraction, CVE matching for estimation, and Reddit classification. SoC/SDK lineage inference is automated but heuristic, with manual review of final attributions. Manual checks cover extraction edge cases, positive CVE detections, Reddit ground-truth labeling, and qualitative analysis for community and regulation portions. A detailed diagram capturing the entire flow is given in Figure-~\ref{fig:soho-pipeline-full}

\end{document}

%% file: abstract.tex
\noindent Small Office/Home Office (SOHO) devices are widely popular, yet often attacked due to security vulnerabilities in their firmware, affecting tens of thousands of devices at a time. These security vulnerabilities often stem from  outdated Linux kernel versions included in SOHO device firmware. Naturally, prior work audited the extent and impact of this issue by simple Linux version extraction and version number based vulnerability mapping. 
However, it is unclear how many of these anticipated vulnerabilities actually exist in the heavily customized SOHO kernels and if there are any barriers towards updating Linux kernels in SOHO firmware. 

To address this gap, we uncover actual kernel-related vulnerabilities found in 306 SOHO devices using a high-precision template-based CVE detection mechanism on GPL source releases of more than 900 firmwares from these devices (multiple versions per device). %
Next, as a first, we traced the supply chain of these vulnerable SOHO devices at scale and identify \textit{kernel lock-in} as a significant security issue---SOHO vendors are effectively locked to specific (often older) kernel versions due to the system-on-chip (SoC) SDKs they use. This kernel lock-in produces a vulnerability debt that is inherited along the supply chain from SoC vendor to firmware creators (ODM/OEM) to router/IP-camera vendor and ultimately borne by end users. All five SoC vendors in our dataset have used SDKs with Linux kernels that had reached End of Life (EoL) more than a  year before their usage in a SOHO device. Finally, we explore the mitigation-potential of individual, regulatory and community governance by analyzing social media posts, regulations and community efforts. Our results show that regulation compliance is insufficient and only SoC vendors who engage with communities for kernel upgradation offered a viable path towards mitigation. 
We conclude by discussing broader implications of our work for improving holistic security of the supply chain for SOHO devices. The data and code for this work is available at \url{https://doi.org/10.5281/zenodo.20433799}

%% file: introduction.tex
\textbf{S}mall \textbf{O}ffice/\textbf{H}ome \textbf{O}ffice (SOHO) devices like routers or IP cameras are characterized by their low cost, support for basic functionality, ease of deployment, and minimal maintenance overhead. However, these devices have also emerged as a significant security threat due to their sensitive position in the network edge. A successful compromise of a SOHO device can enable adversaries to monitor, redirect, or manipulate traffic, pivot into internal hosts, and conscript the device into broader malicious infrastructure. The combination of high privilege, weak access control (e.g., unsafe default configurations and exposed management services), long and unsupported lifecycles, and infrequent firmware updates significantly increases the risk of exploitation \cite{talos_vpnfilter_2018,jsca_prc_linked_botnet_2024,securityscorecard_volt_typhoon_2024,doj_court_operation_botnet_2024, tp_link_quad7_2025}.

In fact, recent exploitations of SOHO devices  are often large-scale, infecting hundreds of thousands of devices~\cite{bleepingcomp_raptor_train_2024} over a long period of time~\cite{fortinet_shadowv2_2025}. Prior work identified the Linux kernel component of SOHO firmware as one of the significant key sources for reported vulnerabilities~\cite{10.5555/3620237.3620429, 10.1145/3533767.3534366, weidenbach-2020}. Moreover, the kernel versions which reached their end-of-life and stopped receiving active security patches are more vulnerable than actively maintained ones. Thus, perhaps not surprisingly,  attacks often exploit SOHO devices with older software/hardware stacks -- many  that have reached the end-of-life~\cite{securityweek-dlink-discontinued-2026, fbi-eol-proxies-2025, cisa_aa22_054a, securityscorecard_wrthug_2025}. 

To understand the true susceptibility on SOHO devices we needed to estimate the vulnerabilities \textit{actually present} in firmware kernels. Prior work simply leverage CVE and kernel version (extracted from firmware binary) mapping to enumerate vulnerabilities (potential overestimation) or use dynamic analysis to confirm the existence of a particular CVE (potential underestimation) ~\cite{10.1145/3427228.3427294, ChenWBE16, 10.5555/3620237.3620518, weidenbach-2020}. However, there is an important catch: SOHO devices use a stripped down Linux kernel version and recent works has hinted that those stripped down kernel versions might not even be chosen by SOHO device vendors~\cite{10.1145/3617072.3617110}. Rather there is a \textit{supply chain} of multiple actors which creates the SOHO devices. As a result there might even be a \textit{kernel lock-in} where in the upstream, one actor might decide on using a particular (often older) Linux kernel version which the downstream actors do not change, resulting in a SOHO device deployed in the wild with potentially vulnerable firmware. 

Unfortunately, there is no prior research, which tried to ascertain the actual vulnerabilities in deployed SOHO device kernels at scale, trace back the origin of those vulnerabilities to potential kernel lock-in along the supply chain, and investigated the efficacy of current holistic mitigation techniques which might address the supply chain issue. In this work we bridge this gap by specifically performing data-grounded exploration of four key research questions.

\textit{{\textbf{RQ1.}} What vulnerabilities are actually present in shipped SOHO device firmware?}

We collect the 977 firmware binaries (859 firmware source codes) of 306 SOHO devices drawn from 5 major vendors. The \textit{source codes} were released under the GNU General Public License (GPL). Using them together, we develop a systematic, code-centric approach correlating firmware images with vendor GPL source trees and kernel build artifacts (e.g., Kconfig and makefile evidence) to infer which Linux kernel CVEs are actually present on each device. We found almost 99\% less CVEs than prior works which extracted Linux Kernel version from binaries and mapped those versions to CVEs. 

\textit{{ \textbf{RQ2}.} How these vulnerabilities are introduced in the supply chain of SOHO devices?}

We uncovered that SOHO router firmware passes through multiple hands: chipset makers, device manufacturers, and the branded vendor. We created a pipeline to map each firmware to its device vendor (along with the OEM/ODM contribution from source code), the System on Chip (SoC) vendor and the Software Development Kit (SDK) that SoC vendors provide (containing stripped down linux kernel). Using this pipeline we show that 94.7\% of SOHO devices in our corpus use SoCs created by only five upstream SoC vendors, and 27.5\%(11 out of  40 SoC SDK kernels) rely on Linux kernel lines that were already end-of-life at the time of deployment.

\textit{{ \textbf{RQ3}.} Do non-SoC actors in the supply chain make a secure kernel selection to resolve the kernel lock-in vulnerability?}

We find that downstream actors usually keep the Linux kernel version inherited from the SoC vendor's SDK. We refer to this inherited SDK kernel as the kernel baseline. Across related device models and hardware revisions, vendors generally retain this baseline instead of moving to a newer, supported kernel. This suggests that kernel upgrades are not treated as a priority in the firmware-integration pipeline, even when non-SoC actors could have selected another kernel.

\textit{{ \textbf{RQ4}.}  Can individual concerns, community efforts or regulator compliance address the Kernel lock-in issue in the supply chain of SOHO devices?}

We used an ML-driven pipeline to systematically detect and investigate 4,104 Reddit posts which mentioned security and privacy issues of SOHO devices. We also checked core regulations (like ETSI EN 303 645~\cite{etsi303645}, NIST IR 8259~\cite{nistir8259})  governing the security ecosystem of SOHO devices, which all organizations of the supply chain should abide by. However, we found that neither end users nor regulators are explicitly addressing kernel lock-in today. Finally, we show that community efforts (such as OpenWrt~\cite{OpenWrt}) in collaboration with SoC-vendors potentially mitigate the kernel lock-in. However, relatively few SOHO devices benefit due to uneven device coverage and their proprietary drivers.

%% file: background.tex
Small Office/Home Office (SOHO) devices (e.g., routers and IP cameras) are consumer-grade embedded networked systems with components like embedded hardware and vendor-customized firmware. %

\noindent
\subsection{SOHO hardware and software stack}

Each SOHO device hardware contains a low-power System-on-Chip (SoC) (e.g., ARM or MIPS), on-board RAM, flash, and network interfaces (e.g., Ethernet switch, Wi-Fi chipset) onto a single Printed Circuit Board (PCB)~\cite{s23229221}. In order to facilitate creating a software stack for this hardware, SoC designers commonly take a mainstream Linux release and build only a subset of its components required for the given SoC. They bundle this stripped-down Linux kernel with custom device drivers and create a Software Development Kit (SDK). These SDKs, including build systems, configuration templates, and platform-specific code for the target SoC family, are then used as integration starting points by SOHO manufacturers. Thus, SOHO software stacks are mostly vendor-customized firmware, often based on Linux kernel~\cite{weidenbach-2020, 10.1145/3533767.3534366}. %

\subsection{Multi-organization supply chain for SOHO}\label{sec:background_supply_chain}

The hardware and software stacks of SOHO devices typically involve a multi-organization supply chain as follows.

\vspace{1mm}
\noindent 
\textbf{SoC vendors}: SoC vendors (e.g., MediaTek ~\cite{mediatek-website}, Qualcomm ~\cite{qualcomm-website}) create the embedded hardware, provide SDK and reference designs. These SDKs, as mentioned, are often built on top of a stripped-down version of Linux kernels. 

\vspace{1mm}
\noindent 
\textbf{Device vendors and OEM/ODM}: Next, device vendors (e.g., D-Link ~\cite{dlinkWebsite}, TP-Link ~\cite{tplinkWebsite}) outsource the development of the firmware based on these SDKs to their OEM (Original Equipment Manufacturer) and ODM (Original Design Manufacturer), e.g., Foxconn ~\cite{foxconn-website}. At the end, the device vendors ship the resulting firmware images. Naturally, many firmware (even from different device vendors) are built for the same SoC family, use the same/similar SDKs, and largely reuse codebase~\cite{10.1145/3617072.3617110, 10.1145/3533767.3534366}. Since OEM/ODMs are employed by device vendors, we consider them as a single entity. 

\vspace{1mm}
\noindent 
\textbf{Open-source firmware community}: SOHO ecosystem also includes open-source firmware projects (e.g., OpenWrt~\cite{OpenWrt}) and developer communities. They maintain alternative build systems, drivers, and device-specific knowledge, often providing alternate firmware with additional features.

\vspace{1mm}
\noindent 
\textbf{Regulators}: Government regulators and compliance bodies (e.g., FCC~\cite{fcc_equip_auth}) act as external stakeholders by shaping the requirements vendors must satisfy (e.g., update support, vulnerability handling). They influence development strategies and security policies for firmware.

\subsection{Kernel Lock-in in SOHO supply chain}

\noindent We use \textit{Kernel Lock-in} to describe a poor supply-chain choice in which a SoC vendor includes in SDK, or a device vendor continues to use, an older Linux kernel instead of moving to a newer supported kernel. Devices affected by kernel lock-in may therefore ship firmware based on older kernel versions.  When such kernels lack upstream security fixes, known vulnerabilities can propagate to all SOHO devices whose firmware is built on the same kernel, creating what we call \textit{Vulnerability Debt}. However, it is not clear where kernel lock-in is initiated in the supply chain, how it propagates across SOHO devices, what the security impact is, or whether any community or regulatory efforts effectively mitigate this issue. In this work, we take a data-driven view, utilizing shipped firmware artifacts, device vendor-released firmware sources, and platform lineage (SoC and SDK) analysis to answer these crucial questions for the first time.

%% file: related_work.tex
We review related work across three broad dimensions of SOHO security: user and community perceptions, security auditing, and CVE detection in SOHO device kernels.

\vspace{1mm}
\noindent
\textbf{User and community perceptions of SOHO security}: Prior work leveraged app reviews, forums, and social media posts (e.g., Reddit), to check the user perceptions of security and privacy for software-centric environments (e.g., mobile apps, or online services). They found that only a small fraction of the user-generated content discusses such concerns~\cite{nguyen2019short, raj2024just,akgul2024decade,mukherjee2020empirical}. Furthermore, these security and privacy concerns do not align with actual behavior, which prioritizes functionality~\cite{Kumaraguru_Cranor:2005, 10.1145/2858036.2858214, 10.1016/j.tele.2017.04.013}. However, it remains unclear if these findings are applicable to hardware-centric systems like IoT devices. Prior work, which checked IoT vulnerabilities, rarely investigated how users discuss security concerns in real-world settings for specific devices~\cite{nino2024unveiling, antonakakis2017understanding, fernandes2016security, adam2024survey, deep2022survey, vetrivel2023examining}. Aside from individual user perception, there is also not much work on uncovering the adoption of community-driven efforts like open-source firmware projects such as OpenWrt~\cite{OpenWrt,codelinaro2026qsdk} or the impact of regulatory and governance frameworks for IoT security~\cite{fcc_equip_auth}. In contrast, we take a holistic view regarding perceptions for SOHO security utilizing online user discussions, community-driven efforts, and regulatory expectations.

\vspace{1mm}
\noindent
\textbf{Auditing SOHO devices for security issues}: Although prior work did not explicitly check user and community perception regarding SOHO device security, a large number of prior studies have characterized SOHO device firmware as vulnerable. Early large-scale firmware analysis showed that static analysis across many embedded firmware images can uncover vulnerabilities and reveal reuse across otherwise unrelated products~\cite{10.5555/2671225.2671232}. They show that Linux is the dominant operating system in such devices and that firmware stacks integrate extensive third-party components ~\cite{weidenbach-2020,10.1145/3533767.3534366}. Some prior works even highlighted how Third-Party Component (TPC) usage patterns can contribute to vulnerabilities and even emulated whole firmware to perform dynamic analysis to find vulnerabilities~\cite{10.1145/3533767.3534366,10.5555/3620237.3620429,ChenWBE16,10.1145/3427228.3427294,9152796}. These works are complementary to our research---we focused on kernel-centric vulnerabilities in SOHO firmware. Next, we discuss prior research on detecting CVEs in SOHO kernels, as we build on those works.

\vspace{1mm}
\noindent
\textbf{CVE detection in SOHO kernel: } 
Earlier works, at their core, estimated kernel vulnerabilities using kernel version-string centric CVE attribution~\cite{weidenbach-2020, 10.1007/978-3-031-35504-2_10}. They either simply looked up all CVEs that correspond to a kernel version string or considered a filtered-down version, considering configuration and build-context constraints inferred from a firmware binary. However, since SOHO firmware often uses stripped-down kernels, these approaches (which do not consider firmware source code) often overestimate kernel security vulnerabilities. To that end, we develop a pipeline to obtain a more precise estimate of kernel vulnerabilities (which are \textit{actually} present in SOHO device firmware) using a code template-based approach applied on source code to detect known CVEs (independent of kernel version). This estimate was crucial to rigorously answer our research question. %

Next, we start by first collecting SOHO firmware binaries, their source code, and tracing down their supply chain. %

%% file: data_collection.tex
In this section, we detail the collection of our primary dataset---SOHO firmware binaries and source code. Furthermore, we collected firmware metadata, e.g., device model identifiers, firmware version strings, device release dates, and End-of-Life (EoL) information from vendor support pages/portals.

\subsection{Collecting SOHO firmware at scale}

\vspace{1mm}
\noindent
\textbf{Creating a SOHO firmware binary corpus}: We start with the firmware dataset from prior work by Kim et al. ~\cite{10.1145/3427228.3427294}, which contains 1141 device firmware from eight vendors. Since our approach requires firmware binaries, corresponding source releases, and device metadata, we restrict our analysis to 306 devices from five popular SOHO vendors who released firmware source code tarballs for at least some devices under the GNU General Public License  (GPL): D-Link~\cite{dlinkWebsite}, TP-Link~\cite{tplinkWebsite}, Linksys~\cite{linksysWebsite}, NETGEAR~\cite{netgearWebsite}, TRENDnet~\cite{trendnetWebsite}.
\begin{table}[]
\centering
\scriptsize
\caption{Number of GPL source code tarballs downloaded for each vendor from their official support pages.}
\begin{tabular}{|r|r|}
\hline
\rowcolor[HTML]{C0C0C0} 
\textbf{Vendor} & \textbf{\begin{tabular}[r]{@{}r@{}}\# GPL source tarballs \\ downloaded\end{tabular}} \\ \hline

D-Link           & 476                    \\ \hline
TP-Link          & 109                   \\ \hline
TRENDnet        & 164                   \\ \hline
NETGEAR         & 26                    \\ \hline
Linksys         & 84                    \\ \hline
\textbf{Total}  & \textbf{859}          \\ \hline
\end{tabular}

\label{table-count-GPL}
\end{table}

\vspace{1mm}
\noindent
\textbf{Identifying SOHO devices from firmware and creating firmware source code corpus using GPL tarballs}: We extracted the device model names for 977 firmware in Kim et al.'s ~\cite{10.1145/3427228.3427294} dataset since the firmware names often contain the names of devices (e.g., \texttt{DIR820LA1\_FW100KRB05}). We searched these device names on official vendor portals to download the latest available firmware binaries for these devices and ended up with 1133 firmware binaries. Next, in order to detect the source codes, we leverage a simple idea---many of the device vendors release their source code tarballs under the GPL. Thus, we created a Selenium browser automation-based pipeline~\cite{gundecha2018selenium}. Our pipeline simply visited each of the five vendors' portals (these public links were manually curated), meant to provide firmware source code for different device models, and downloaded the source tarballs. We apply conservative wait times to ensure tarball retrieval while putting minimum overhead on these portals. When available, we also collected portal metadata for firmware source tarballs (e.g., posting dates or release notes). In total, we extracted 859 unique firmware tarballs for 1133 firmware binaries across 306 devices from five vendors. Notably, some unique firmware tarballs corresponded to multiple firmware binaries, potentially since those binaries used the same core codebase with very minor differences (e.g., enforced at the UI level). Furthermore, some device model names had multiple firmware source tarballs associated with them. We considered only the tarballs with the latest kernel version. Table~\ref{table-count-GPL} presents the breakdown of 859 tarballs across five vendors. Next, we trace the organizations in the supply chain that created these firmware.

\begin{table}[]
\centering
\scriptsize
\caption{Number of SoCs identified in devices for each vendor through public sources and firmware source tree.}
\begin{tabular}{|r|r|r|r|}
\hline
\rowcolor[HTML]{C0C0C0} 
\textbf{Vendor} & \textbf{\# Devices} & \textbf{\begin{tabular}[r]{@{}r@{}}\# Devices w/ \\   identified SoC\end{tabular}}  & \textbf{ \begin{tabular}[r]{@{}r@{}} \% of Devices w/ \\identified SoCs\end{tabular}} \\ \hline
D-Link & 64 & 58 & 90.6 \\ \hline
TP-Link & 104 & 86 & 82.7 \\ \hline
TRENDnet & 54 & 46 & 85.2 \\ \hline
NETGEAR & 29 & 26 & 89.6 \\ \hline
Linksys & 55 & 49 & 89.0 \\ \hline
\textbf{Total} & \textbf{306} & \textbf{265} & \textbf{86.6} \\ \hline
\end{tabular}

\label{table-soc-covered}
\end{table}

\subsection{Discovery of supply chain for SOHO firmware}

\noindent Recall from Section~\ref{sec:background_supply_chain} that the creation of a hardware-software stack of SOHO devices involved a multi-organizational supply chain. In this section, starting from the firmware binary/firmware tarball, we seek to trace the organizations in the supply chain of each device model.

\begin{figure}
    \centering
\includegraphics[width=1\linewidth]{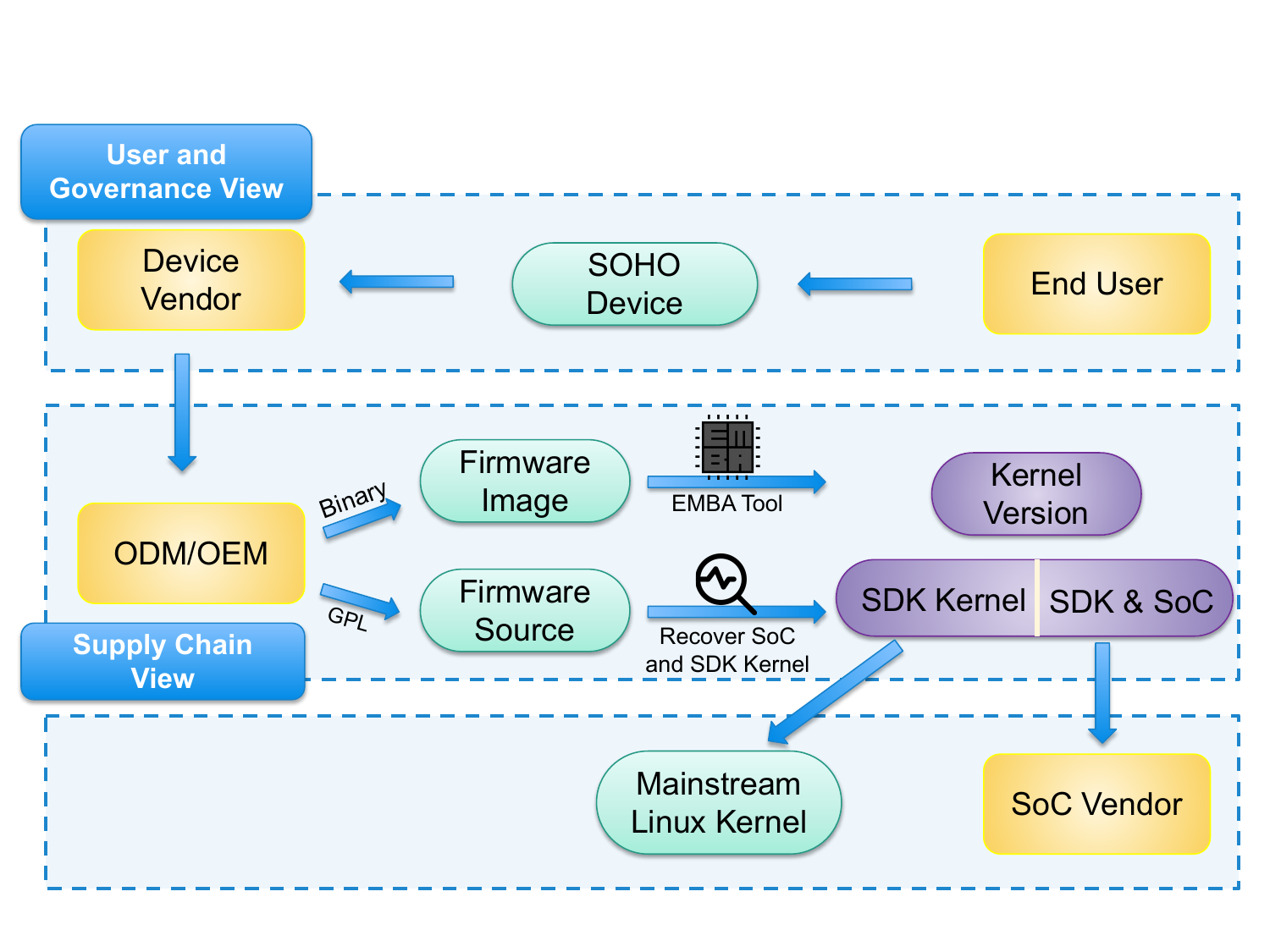}
    \caption{Discovery of the Supply Chain via SOHO device firmware analysis.}
    \label{supply-chain-flowchart}
\end{figure}

We take a systematic top-down approach to trace the supply chain---presented in Figure~\ref{supply-chain-flowchart}. In particular, we use firmware tarballs to infer the SoC family and the SDK kernel baseline, and relate these upstream baselines to the kernel that ultimately appears in shipped firmware. To our knowledge, this is the first SOHO-focused supply-chain discovery approach that correlates vendor firmware binaries, GPL source releases, SoC evidence, and SDK lineage at scale. We envision this \emph{supply chain discovery} step as a prerequisite for future supply-chain-grounded analysis. In the rest of the paper, we will consider devices with complete supply chain tracing, unless the analysis does not require SoC/SDK attribution. Note that Figure~\ref{supply-chain-flowchart} is a simplified end to end view of our pipeline. We discussed the exact technical details of various part of the pipleine as well as relevant results throughout the paper. For interested readers, we present a version of our pipeline with additional information (e.g., analysis is automated or manual, exact techniques used) in Figure~\ref{fig:soho-pipeline-full} of Appendix~\ref{app-large-fig}.

\vspace{1mm}
\noindent \textbf{SoC identification:} We map the device firmware to a specific SoC family using a tiered approach. We first infer the SoC identity from the build evidence in the vendor’s firmware source release by extracting SoC-relevant configuration and build-flag hints. Concretely, we search for candidate kernel configuration files in the source tree and parse enabled \texttt{CONFIG\_} symbols (\texttt{=y} or \texttt{=m}), retaining only those likely to encode SoC, board, vendor, or architecture identity (e.g., prefixes such as \texttt{SOC\_}, \texttt{ARCH\_}, \texttt{MACH\_}, \texttt{BOARD\_}, \texttt{VENDOR\_}, as well as vendor/SoC-family keywords such as \texttt{IPQ}, \texttt{MTK}, \texttt{BCM}, \texttt{RTL}). We then rank these symbols by signal strength to extract a compact set of high-confidence SoC indicators and use them to infer the SoC model when an exact identifier is present; when only family-level evidence is available, we retain the SoC family as a hint and treat the SoC label as provisional. We then consult public sources (e.g., Wikidevi~\cite{WikiDevi}, TechInfoDepot~\cite{TechInfoDepot}, and OpenWrt~\cite{OpenWrt} forums) to obtain SoC information when it is explicitly documented and to corroborate or refine source-derived SoC hints when possible. In cases where both public sources and source-derived configuration evidence are available, we prefer exact SoC model names from the firmware source and use family-level hints only when exact model identifiers are missing. Table~\ref{table-soc-covered} presents the coverage per vendor.

\vspace{1mm}
\noindent

\vspace{1mm}
\noindent
\textbf{SDK discovery from firmware source releases}: We infer the SDK lineage for a device using two complementary sources of evidence. First, we analyze vendor firmware source releases to locate SDK repositories in the source tree and use corroborating SDK markers such as SDK-specific naming conventions, SDK-specific Makefiles, SDK documentation artifacts (e.g., README files), and other build evidence—before classifying an SDK lineage as present. Second, when available, we record SDK information from public repositories and community sources (e.g., Wikidevi, TechInfoDepot, and OpenWrt forums) that explicitly name SDK, and we use this public attribution to corroborate or refine the source-derived SDK labels. When the same SoC appears across multiple device models (and, in some cases, across vendors), we use consistency of these SDK markers across the corresponding source releases as an additional sanity check on SDK identification. We present summary statistics of our final SDK verification coverage in Table~\ref{table-sdk-covered}; we treat one SDK per SoC as one unit and count an SDK as verified only when the firmware source release contains sufficient SDK markers and build artifacts to attribute an SDK lineage with confidence.

\vspace{1mm}
\noindent
\textbf{Extracting the SDK kernel baseline and validating consistency with firmware binary: }
Next, we extract the Linux kernel baseline associated with that SDK portion of the source release by analyzing the Linux kernel Makefiles in each of the source trees in an X.Y.Z (e.g, \texttt{VERSION=2}, \texttt{PATCHLEVEL=6}, \texttt{SUBLEVEL=36}) format. We then treat this SDK-derived kernel baseline as the baseline for the device and verify that it is consistent with the kernel baseline observed in the corresponding firmware binary when kernel artifacts are available (more details are in Appendix~\ref{app_sec4}). We leverage this data to answer RQ1: measuring actually present CVEs in firmware kernels.

\begin{table}[]
\centering
\scriptsize
\caption{Number of SDKs identified for each vendor.}
\begin{tabular}{|r|r|r|r|}
\hline
\rowcolor[HTML]{C0C0C0} 
\textbf{Vendor} & \textbf{\# Devices} & \textbf{\begin{tabular}[r]{@{}r@{}}\# SDK \\ identified\end{tabular}} & \textbf{\begin{tabular}[r]{@{}r@{}}\% SDK \\ identified\end{tabular}} \\ \hline

D-Link           & 64   & 56    & 87.5       \\ \hline
TP-Link          & 104     & 77      & 74.0           \\ \hline
TRENDnet        & 54   & 42     & 77.7             \\ \hline
NETGEAR         & 29      & 27   & 93.1            \\ \hline
Linksys         & 55     & 41        & 74.5         \\ \hline
\textbf{Total}  & \textbf{306}  & \textbf{243}   & \textbf{79.4}     \\ \hline
\end{tabular}

\label{table-sdk-covered}
\end{table}

%% file: RQ/newRQ1.tex
In this section, we first assemble a firmware-source analysis pipeline to infer CVEs actually present in deployed SOHO firmware and then compare the number of detected actual CVEs with the version-string based CVE estimate used in prior work.

\subsection{Our pipeline for detecting CVEs}

\noindent
\textbf{Firmware source-based kernel extraction and template-based CVE inference}: We first perform firmware source-based kernel analysis to infer vulnerability presence from the shipped kernel source tree and its build context. For each firmware source tarball, we locate the Linux kernel subtree by using multiple signals: presence of canonical kernel markers (e.g., Kconfig and kernel Makefiles), directory structure checks consistent with Linux source layout, and keyword-based cues (e.g., “linux”) within the extracted tree. We use Coccinelle~\cite{coccinelle_website} as the template-matching tool in our pipeline: it is a widely used semantic search and program transformation tool for kernel source code analysis. We automate template-based matching by using cvehound~\cite{cvehound-Github}, which leverages Coccinelle and its \texttt{spatch} engine under the hood together with a public collection of CVE-specific semantic templates, allowing us to systematically scan vendor kernel source trees for known vulnerability patterns at scale. In the presence of multiple firmware source tarballs for a single device model, we run our analysis for all available firmware sources, but we report results using the latest firmware source release with the latest Linux kernel baseline. %

\vspace{1mm}
\noindent
\textbf{Patch-template checks (supporting evidence only)}: In our pipeline, we also deployed a limited fix-oriented check for a subset of cases. Once a vulnerable code pattern is matched in a kernel tree via CVE templates, we additionally run the corresponding “fixes” (patch) template on the same tree when an upstream fixing pattern is available. The combination of a positive vulnerable-pattern match and the absence of the corresponding fixes-pattern match provides a precise signal that the vulnerable code remains present without the known upstream fix, and thus strengthens our per-device inference of which CVEs are present in the shipped kernel.

\vspace{1mm}
\noindent
\textbf{Baseline: Kernel version-string based CVE estimation in firmwares}: We leveraged the firmware binary-centric kernel baseline extraction used in prior large-scale firmware measurements ~\cite{weidenbach-2020, 10.1007/978-3-031-35504-2_10}. Specifically, for each firmware binary, we unpack and analyze the binary using the EMBA~\cite{embaGitHub} tool suite and extract the Linux kernel version using linux-exploit-suggester~\cite{thez-labs_linux-exploit-suggester} (more details are in Appendix~\ref{sec:binary_analysis_baseline}). Next, we simply map the kernel version to the known CVEs, following earlier work as our version-string baseline.

\begin{table}[]
\centering
\scriptsize
\setlength{\tabcolsep}{4pt}
\caption{Comparison of the number of unique CVEs found via version-string centric attribution and build-aware firmware source CVE inference, and the percentage decrease in inferred CVE counts per device (averaged per vendor). }
\begin{tabular}{|r|r|r|r|}
\hline
\rowcolor[HTML]{C0C0C0} 
\textbf{Vendor} & \textbf{\begin{tabular}[r]{@{}r@{}} Baseline: Version-string \\ CVE attribution\end{tabular}} & \textbf{\begin{tabular}[r]{@{}r@{}}Our firmware source\\  CVE attribution\end{tabular}} & \textbf{\begin{tabular}[r]{@{}r@{}}\% Decrease\\ in \# CVE\end{tabular}} \\ \hline
D-Link & 4249 & 43 & 99.71  \\ \hline
TP-Link & 3884 & 22 & 99.66  \\ \hline
TRENDnet & 4318 & 23 & 99.91  \\ \hline
NETGEAR & 4454 & 26 & 99.73  \\ \hline
Linksys & 3732 & 22 & 99.90  \\ \hline
\end{tabular}

\label{table-cve-percent2}
\end{table}

\subsection{Results}

\textbf{Significant over-estimation of CVEs in SOHO firmware kernels in prior work}: Table~\ref{table-cve-percent2} presents the number of CVEs attributed to each firmware in our dataset using the version string (following prior work) and our firmware source base CVE attribution pipeline. Strikingly, on average, across five vendors, we observed a 99.8\% decrease in the number of CVEs when we enforced source code-based attribution (designed to detect CVEs actually present in the firmware). 

\vspace{1mm}
\noindent
\textbf{Investigating potential reason for over-estimation}: We next check the actual CVE IDs to better understand this difference between earlier kernel version-centric attribution and our build context-aware template-based inference. Table~\ref{tab:top-cves-version-vs-gpl} contrasts the top-5 CVEs identified via version-string matching with the top-5 CVEs inferred using our approach, 
We observe that in the version-string-centric approach, the ``top CVEs'' are dominated by vulnerabilities in subsystems that are plausible to be absent from a lean SOHO kernel build (e.g., Open vSwitch in CVE-2024-1151~\cite{nvd:CVE-2024-1151}, or device-specific graphics drivers in CVE-2024-22386~\cite{nvd:CVE-2024-22386}). In contrast, our approach surfaces issues in core IPv4 networking components that are more plausibly enabled in router kernels (e.g., IGMP handling in CVE-2023-6932~\cite{nvd:CVE-2023-6932}). Thus, a potential reason for the overestimation was to simply consider what is possible for a general kernel version, not what is actually instantiated in a shipped kernel tree.

%% file: RQ/newRQ2.tex
Our pipeline identified a non-trivial (although decreased) number of confirmed vulnerabilities actually present in deployed SOHO firmware, which is quite concerning. Thus, next in RQ2, we trace where and why this kernel vulnerability debt is introduced in the supply chain. 

\begin{table}[t]
  \centering
  \scriptsize
  \caption{Top-5 kernel CVEs under version-string attribution versus firmware source inference.}
  \begin{tabular}{|rr|rr|}
\hline
\rowcolor[HTML]{C0C0C0} 
\rowcolor[HTML]{C0C0C0} 
\multicolumn{1}{|r|}{\cellcolor[HTML]{C0C0C0}\textbf{Rank}}           & \textbf{\begin{tabular}[r]{@{}r@{}} Version-string \\ CVE attribution\end{tabular}}             & \multicolumn{1}{r|}{\cellcolor[HTML]{C0C0C0}\textbf{Rank}}          & \textbf{\begin{tabular}[r]{@{}r@{}}Our Firmware source\\  CVE attribution\end{tabular}}                \\ \hline
\multicolumn{1}{|r|}{1} & CVE-2024-0607  & \multicolumn{1}{r|}{1}    & CVE-2014-7975 \\ \hline

\multicolumn{1}{|r|}{2}                                      & CVE-2024-0641          & \multicolumn{1}{r|}{2}                                     & CVE-2020-29661          \\ \hline

\multicolumn{1}{|r|}{3}                                      & CVE-2024-1151          & \multicolumn{1}{r|}{3}                                     & CVE-2023-6932          \\ \hline

\multicolumn{1}{|r|}{4}                                      & CVE-2024-1312          & \multicolumn{1}{r|}{4}                                     & CVE-2024-26733         \\ \hline

\multicolumn{1}{|r|}{5} & CVE-2024-22386          & \multicolumn{1}{r|}{5}                                     & CVE-2020-10732         \\ \hline
\end{tabular}

  \label{tab:top-cves-version-vs-gpl}
\end{table}

\subsection{Methodology}
\noindent
\textbf{Detecting introduction of kernel vulnerabilities and kernel lock-in in supply chain}: We treat the SoC-vendor SDK as the earliest observable point where a Linux kernel version is selected and then reused downstream. We call this SDK-provided kernel version the SDK \textit{kernel baseline}. Concretely, we detect ``SDK presence'' by requiring corroborating SDK markers observed consistently across a fixed set of ``interesting'' lineage files (e.g., Makefile, README variants, RELEASE/version files). SDK markers include SDK/Board Support Package (BSP) related naming, SDK-specific build artifacts and Makefiles, SDK documentation artifacts, and SoC-family identifiers; we treat the presence of multiple markers as stronger evidence, and we assign higher preference to direct lineage indicators (e.g., OpenWrt-derived Makefiles) when they appear. We cap scanning to the first 400k characters per file to ensure broad coverage while bounding noise. We treat a firmware source tree as SDK-linked only when we observe consistent SDK markers such as SDK or BSP-related naming, SDK-specific build artifacts, or SoC-family identifiers across these files. We then extract the \textit{kernel baseline} from the SDK-linked portion of the tree and normalize it to a three-component version (e.g., 2.6.36) for comparison. To test whether this SDK kernel baseline becomes locked-in downstream, we compare it to the Linux kernel version extracted from shipped firmware binaries and treat matches (to three components) as evidence that the shipped kernel follows the same upstream kernel track.

\vspace{1mm}
\noindent
\textbf{Using kernel baseline reuse as lock-in evidence}: We evaluate whether downstream firmware deviates from the SDK kernel track by operationalizing “inheritance” as a version match between (i) the SDK Linux kernel version extracted from firmware source release and (ii) the Linux kernel version extracted from the corresponding firmware binary. Concretely, for each device where both signals are available, we extract and normalize both kernel versions to three version components (e.g., 2.6.36) and treat a match as evidence that the shipped kernel follows the same SDK provided kernel version. We then aggregate these matches across devices and stratify them by SoC and vendor to enable downstream analysis of reuse patterns, including cases where the same SoC appears across multiple device models and across vendors.

\vspace{1mm}
\noindent
\textbf{Detecting EoL for locked-in kernels at time of use}: To characterize when locked-in SDK kernel baselines are already outside upstream support, we compare the SDK kernel baseline against official Linux kernel end-of-life timelines, using the SoC release year as a conservative proxy for when that SoC's SDK would have been introduced. For SoCs whose SDK kernel line is observed across multiple devices (i.e., shared SDK kernel lines), we recover SoC release years from public sources and use these as stable anchors for “time of platform introduction”. While the SDK for a SoC may be released later than the SoC itself, using the SoC release year biases our analysis toward underestimating the extent of the problem rather than overstating it. We flag an SDK kernel baseline as problematic when the SoC release year is the same as or later than the kernel’s end-of-life year, since this indicates the SDK is built on a kernel line for which no upstream-maintained patch stream exists, sharply limiting downstream stakeholders' ability to inherit fixes through mainstream maintenance.

\subsection{Results}

\begin{table}[]
\centering
\scriptsize
\caption{Representation of each major SoC-vendor in the identified list of SoCs.}
\begin{tabular}{|r|r|}
\hline
\rowcolor[HTML]{C0C0C0} 
\textbf{SoC Vendor} & \textbf{{\begin{tabular}[r]{@{}r@{}}\% of SoCs in \\identified list\end{tabular}} } \\ \hline
MediaTek            & 30.1               \\ \hline
Broadcom            & 26.0               \\ \hline
Qualcomm             & 19.6               \\ \hline
Realtek             & 11.7               \\ \hline
Atheros             & 7.1               \\ \hline
\textbf{Total}      & \textbf{94.7}               \\ \hline
\end{tabular}

\label{tab-soc-vendor}
\end{table}

We analyzed the supply chains for all the firmware in our dataset and made two interesting observations. 

\vspace{1mm}
\noindent
\textbf{Multiple SOHO devices reuse kernel due to reuse of the same SoC/SDK from a small number of vendors}:
First, across our dataset, multiple devices share common SoCs (a total of 102 SoCs across 306 SOHO devices); we identify 52 SoCs that are used by more than one device and, in total, cover 70\% of devices. Overall, 94.7\% of SoCs used in SOHO devices from our corpus use SoCs created by only five upstream SoC vendors (shown in Table~\ref{tab-soc-vendor}), implying that upstream kernel choices by a handful of SoC vendors can shape the security of many downstream SOHO devices.

\begin{table}[]
\centering
\scriptsize
\caption{Examples of Linux kernel baselines inherited from SoC SDKs, together with kernel End-of-Life (EoL) dates, SoC release timelines, and device release years, illustrating how kernels may end up being shipped onto SOHO devices multiple years after reaching EoL.}
\begin{tabular}{|r|r|r|r|r|}
\hline
\rowcolor[HTML]{C0C0C0} 
\textbf{\begin{tabular}[c]{@{}r@{}}Linux\\ Kernel\\ version\end{tabular}} & \textbf{\begin{tabular}[c]{@{}r@{}}Kernel\\ EOL\\ Year\end{tabular}} & \textbf{\begin{tabular}[c]{@{}r@{}}SoC\\ release\\ Year\end{tabular}} & \textbf{\begin{tabular}[c]{@{}r@{}}Example\\ SoCs\end{tabular}} & \textbf{\begin{tabular}[c]{@{}r@{}}Release year \\ of a device that \\ use this SoC\end{tabular}} \\ \hline
2.6.31                                                                    & 2010                                                                 & 2011                                                                  & AR9341                                                          & 2013                                                                                              \\ \hline
2.6.36                                                                    & 2011                                                                 & 2015                                                                  & BCM4709C0                                                       & 2016                                                                                              \\ \hline
3.10.14                                                                   & 2017                                                                 & 2019                                                                  & MT7628DAN                                                       & 2019                                                                                              \\ \hline
4.4.60                                                                    & 2022                                                                 & 2019                                                                  & BCM6755                                                         & 2020                                                                                              \\ \hline
\end{tabular}

\label{tab-years-lag}
\end{table} 

\vspace{1mm}
\noindent
\textbf{Significant fraction of the reused kernels reached EoL at time of use}: Second, for 40 of these 52 SoCs, we detected if the corresponding SDK kernel version reached end-of-life. Surprisingly, 11 of the 40 SoC SDK kernels rely on Linux kernel lines that were already end-of-life (see Figure~\ref{fig:sdk-kernel-map} in Appendix~\ref{app_kernel_eol}). For 8 of these 11, the SoC release year lags the kernel’s end-of-life by 2 or more years, indicating that SDK baselines are anchored to kernel lines long after upstream support has ended. Table~\ref{tab-years-lag} shows a few representative SoCs. Some of them reached end-of-life even before these SoCs were released. Some SDK baselines use kernels that are upstream supported, while others remain anchored on long-EoL kernel lines. This sets us up for the next question: Did the non-SoC vendor organizations in the supply chain not have a choice other than selecting EoL kernels (RQ3)?

%% file: RQ/newRQ3.tex
In this section, we check where non-SoC vendor organizations in the supply chain can make a choice for the kernel. We identify a crucial integration point---the downstream ODM/OEM developers (controlled by device vendor~\cite{10.1145/3617072.3617110})---where firmware can either remain on the SDK kernel track or detach to a supportable, preferably Long Term Support (LTS), kernel line when building device firmware.

\vspace{1mm}
\noindent
\textbf{ODM/OEM developers can choose to inherit or detach during integration}: Since ODM/OEMs act as integrators, they can either i) ship the SDK kernel as it is, or ii) detach from the SDK kernel and upgrade to an upstream supported kernel (e.g., LTS release) while keeping the same SoC/SDK. We treat detachment as a downstream choice because it is being made during the integration phase and is only visible through what kernel line ends up in the shipped firmware.

\vspace{1mm}
\noindent
\textbf{Choice made by majority ODM/OEM developers}: We compare the version number of SDK-associated kernel line recovered from firmware source releases to the kernel line extracted from shipped firmware binaries. %
 In 174 cases where both are recovered, we find only 12 cases (6.8\%) where the shipped kernel line is more recent than the one found in the SDK-marked region of the firmware source, indicating that shipped firmware rarely upgrades away from the SDK kernel versions during downstream integration as seen in Table~\ref{firmware-gpl-relationship}. We discuss these likely  mismatches in Appendix~\ref{app:firmware-gpl-mismatch}.

\begin{table}[]
\caption{Categorization of firmware kernels based on their version relationship with available firmware source kernels.}
\scriptsize
\centering

\begin{tabular}{|p{5cm}|r|}
\hline
\rowcolor[HTML]{C0C0C0} 
\textbf{Firmware version comparison} & \textbf{Count} \\ \hline
Kernel version in firmware binary < Kernel version in firmware GPL source & 77 \\ \hline
Kernel version in firmware binary > Kernel version in firmware GPL source & 12  \\ \hline
Kernel version in firmware binary = Kernel version in firmware GPL source & 85 \\ \hline
\end{tabular}
\label{firmware-gpl-relationship}

\end{table}

\vspace{1mm}
\noindent
\textbf{Feasibility of kernel upgrade in SOHO firmware (OpenWrt-derived lineages)}: Lastly, we checked if it was indeed feasible for the OEM/ODM developers to upgrade the kernel in the final firmware. We found an exact platform built for this purpose, but leveraged by only a handful of OEM/ODM developers and device vendors. Specifically, we identify devices whose firmware source trees contain OpenWrt source-tree artifacts (e.g., OpenWrt/LEDE structure and associated configuration/README files) and whose SDK strings indicate vendor-declared OpenWrt-derived lineages (e.g., Qualcomm QSDK ~\cite{codelinaro2026qsdk}). One such example from our dataset is the SoC from Qualcomm \verb|IPQ9574|, which uses Qualcomm's QSDK that is built on top of OpenWrt 24.10 and, in turn, uses Linux kernel 6.6, which is the latest kernel supported among all the tested devices.  

Thus, the SOHO supply chain actors largely preserve, rather than consistently mitigate, inherited kernel lock-in in the firmware that ultimately ships. To that end, finally, we explore whether the security preferences of individual users, the community, or government regulations can govern (and mitigate) the security risks associated with kernel lock-in.

%% file: datacollection_RQ4.tex
In this section, we collect large-scale data to quantify governance efforts in the SOHO ecosystem from the perspectives of end users, community-driven firmware projects, and regulatory frameworks. Using this data, we will assess whether these efforts can counter the problem of kernel lock-in.

\begin{table}[]
\scriptsize
\centering
\caption{Performance of our fine-tuned models on the test dataset for classifying Reddit discussions into security-related and non-security-related categories.}
\begin{tabular}{|r|r|r|}
\hline
\rowcolor[HTML]{C0C0C0} 
\textbf{Model}   & \textbf{Accuracy} & \textbf{Macro avg F1} \\ \hline
BERT             & 0.90              & 0.90                  \\ \hline
RoBERTa          & 0.94              & 0.93                  \\ \hline
\textbf{DeBERTa} & \textbf{0.95}     & \textbf{0.95}         \\ \hline
FLAN-T5          & 0.86              & 0.85                  \\ \hline
\end{tabular}

\label{comparative_model_performance}
\end{table}

\subsection{Collecting user perception data }\label{sec_8_1}

We began our analysis with a curated set of 306 consumer devices spanning 5 device vendors and product categories. We use Reddit as a proxy for general-user discourse, in line with prior work~\cite{11023504, li2023s}. We observed that Reddit users refer to devices using a variety of lexical forms, including differences in spacing, hyphenation, and spelling. To account for this variability, we issued multiple (a total of 6 variants) search queries per device. We used browser automation via Selenium~\cite{gundecha2018selenium} and queried Reddit for each device, collecting 4,104 unique posts across 306 devices and their query variants (more details in Appendix~\ref{app_rq4_dc_userperception}). We collected a total of 22,875 unique comments across 4,104 Reddit posts. 
Since security and privacy related discussions are relatively rare in public forums (like Reddit), manually curating a balanced dataset was difficult. Thus, in line with prior work~\cite{li2023s,nguyen2019short}, we applied keyword-based filtering to identify a candidate set of potentially relevant comments before manual labeling. For this purpose, we expand existing keyword lists~\cite{nguyen2019short} through simple dictionary search, but we also note that such filtering remains coarse due to the informal and contextual nature of Reddit discussions.

\subsubsection{Leveraging text classification to identify the security and privacy related comments dataset}\label{sec_8_1_1}

We combined human annotation with supervised machine learning to scale the identification of security and privacy-related comments across our entire dataset.

\vspace{1mm}
\noindent
\textbf{Creating ground truth labels}: We view the problem of identifying security and privacy related discussions within Reddit comments as a binary classification task. To that end, we manually created a gold-standard dataset consisting of 762 manually labelled comments. Two coders independently coded these comments and achieved inter-rater agreement (Cohen’s kappa) of 0.8, indicating substantial agreement. 
Using the manually labeled dataset as ground truth (more details in Appendix~\ref{app_rq4_dc_ml}), we evaluate multiple fine-tuned transformer-based BERT models and select DeBERTa based on its best overall performance, achieving an F1-score of 0.95. Detailed metrics 
are reported in Table~\ref{comparative_model_performance}.

\vspace{1mm}
\noindent
\textbf{Model Performance}: We randomly split the labeled dataset into 80\% training and 20\% validation subsets. We utilized a range of pretrained transformer-based models to address the classification challenge. Specifically, we evaluated BERT~\cite{devlin2018bert}, RoBERTa~\cite{liu2019roberta}, and DeBERTa~\cite{he2020deberta}, FLAN-T5~\cite{chung2024scaling}, all of which were fine-tuned on our labeled dataset for the downstream task of comment classification (details of model architecture given in Table~\ref{details_bertmodels_arch} of Appendix~\ref{app_rq4_dc_ml}). After training, the selected DeBERTa model achieved an accuracy of 95\%, and an F1-score of 95\%. Detailed performance metrics for all evaluated models are reported in Table~\ref{comparative_model_performance}. We further conduct a misclassification analysis, with detailed results reported in Table~\ref{Miscallsification_deberta} (of Appendix~\ref{app_rq4_dc_ml}).

\vspace{1mm}
\noindent
\textbf{Applying the model to scale up security and privacy related Reddit comments}: Finally, we used our validated comment-classification model on the remaining Reddit comments that were not part of the manually labeled dataset. We identified security and privacy related 1,240 comments (5.42\%). This low prevalence suggests that general users remain largely disengaged from the security and privacy challenges underlying SOHO devices, limiting the effectiveness of user-driven governance in this space. %
While Reddit captures general user perception, technical discussions around firmware and kernel-related issues may be more prevalent in dedicated forums. To assess this, we further collected and analyzed posts from OpenWrt, DD-WRT, and vendor forums (e.g., Netgear, TP-Link, SNB); detailed results and are reported in Appendix~\ref{other_than_reddit}.

\subsubsection{Uncovering the topics of security and privacy related Reddit discussions}\label{sec_8_1_2}

Next, we use open coding and affinity diagramming ~\cite{saldana2015coding} to iteratively develop a hierarchy of themes that explain how Reddit users discuss, interpret, and respond to security and privacy related issues in SOHO device related discussions.

\vspace{1mm}
\noindent
\textbf{Open coding}: We first applied open coding to the extracted explanatory quotes. We randomly sampled 248 quotes (which is ~20\% of the total) and used these to collaboratively develop an initial codebook. In addition, we set aside \textit{25} quotes at the outset to check the saturation of themes. Specifically, we started coding 60 ($\sim$5\% of) quotes (in line with previous work ~\cite{raj2024just,ghosh2025wasn}). Subsequently, the two researchers used the codebooks to independently code all the quotes in each phase. Inter-rater agreement (Cohen’s Kappa) at the end of the open coding round was 0.77, indicating substantial agreement. 
At the end of the open coding phase, the researchers met to resolve disagreements and finalize codes, creating new codes when necessary. A held-out set of 25 quotes was then successfully coded using the finalized scheme, indicating thematic saturation~\cite{saunders2018saturation}. In total, the coders assigned 11 distinct codes (L-2 or Level-2 codes).

\vspace{1mm}
\noindent
\textbf{Affinity diagramming to identify the hierarchy of themes}: Following open coding, the researchers collaboratively analyzed the resulting codes using affinity diagramming to group codes into higher-level themes by jointly examining codes and their associated quotes~\cite{10.1145/2702123.2702561}. A subset of random codes (10) was held out to assess saturation, and iterative grouping continued until no new themes emerged, resulting in a 4-level thematic hierarchy (more details in Appendix~\ref{app_rq4_dc_themes}). Level-1 contains abstract highest-level themes created in the final round of affinity diagramming, and Level-4 contains the very specific codes specified by the open coding step. Our hierarchy of themes (first four levels) explaining security and privacy related discussions around SOHO devices is shown in Table~\ref{four_level_hierarchy} (of Appendix~\ref{app_rq4_dc_themes}).

\vspace{1mm}
\noindent
\textbf{Scaling up}: We extend the thematic analysis to our entire dataset using manual annotation, as the labeled data are insufficient for reliable ML model training. Two independent annotators labeled 992 additional comments (of the total 1240) related to security and privacy, achieving an inter-rater agreement of 0.74. Table~\ref{detailed-numbes} summarizes the identified themes.

\begin{table}[]
\scriptsize
\centering
\caption{ The hierarchical distribution of themes identified through manual open coding of Reddit discussions. Hierarchy denotes the level of the theme in the coding structure (L1: top-level, L2: sub-theme). F1 represents the number of quotes in which a theme appeared during manual analysis, while F2 denotes the number of quotes identified while scaling up.}
\begin{tabular}{|
>{\columncolor[HTML]{FFFFFF}}l |
>{\columncolor[HTML]{FFFFFF}}l |l|l|}
\hline
\cellcolor[HTML]{C0C0C0}\textbf{Theme} & \cellcolor[HTML]{C0C0C0}\textbf{Hierarchy} & \cellcolor[HTML]{C0C0C0}F1 & \cellcolor[HTML]{C0C0C0}F2 \\ \hline
\cellcolor[HTML]{CBCEFB}\begin{tabular}[c]{@{}l@{}}A. Related to S/P \\ of soho devices\end{tabular} & \cellcolor[HTML]{CBCEFB}L1 & \cellcolor[HTML]{CBCEFB}55 & \cellcolor[HTML]{CBCEFB}212 \\ \hline
A.1 Configuration issues & L2 & 16 & 43 \\ \hline
A.2 Network security & L2 & 11 & 57 \\ \hline
A3 Awareness & L2 & 9 & 38 \\ \hline
A.4 Exploit & L2 & 5 & 17 \\ \hline
A.5 Update cycle & L2 & 6 & 29 \\ \hline
A.6 Misc & L2 & 8 & 28 \\ \hline
\cellcolor[HTML]{CBCEFB}\begin{tabular}[c]{@{}l@{}}B. Not related to S/P\\ of soho devices\end{tabular} & \cellcolor[HTML]{CBCEFB}L1 & \cellcolor[HTML]{CBCEFB}193 & \cellcolor[HTML]{CBCEFB}780 \\ \hline
B.1 Wifi protocol & L2 & 8 & 52 \\ \hline
B.2 Configuration & L2 & 54 & 275 \\ \hline
B.3 Software related & L2 & 79 & 225 \\ \hline
B.4 Vulnerability of other iot & L2 & 9 & 2 \\ \hline
B.5Misc & L2 & 43 & 226 \\ \hline
\end{tabular}

\label{detailed-numbes}
\end{table}

\subsection{Collecting community data}

Community-driven firmware projects play an important role in extending the security and maintenance lifetime of SOHO devices. While several such communities exist, such as DD-WRT, Tomato, and OpenWrt~\cite{ddwrt, freshtomato, OpenWrt}, we focus on OpenWrt, as it provides the most extensive and systematic public data on device support, kernel versions, and hardware compatibility, and because we find it being used in SDK baselines by vendors. For each device in our dataset, we aim to identify whether the community supports it (or supported it until recently), and if so, the latest OpenWrt version and the corresponding Linux kernel version that OpenWrt provides. To collect this information at scale, we rely primarily on the OpenWrt Table of Hardware (ToH)~\cite{openwrt_toh2026} and archived ToH~\cite{openwrt-github} for recently supported devices, a curated database with device–SoC compatibility, supported OpenWrt versions, hardware specifications, and  reference links. We use a local copy of the toh.json file to automate data extraction.

We matched each device’s vendor, model, and SoC against the ToH entries to obtain the latest supported OpenWrt version and associated supporting references. We then map each OpenWrt release to its corresponding Linux kernel version using OpenWrt release documentation and device-specific pages. For devices not covered by the ToH snapshot, we consult individual OpenWrt device pages and, when needed, upstream source repositories to identify the most recent supported release. Overall, this process yields community support information for 100 devices, which we use in subsequent analyses to characterize the extent to which community governance can sustain kernel support beyond vendor-maintained update lifecycles. Table~\ref{openwrt-support-distribution} summarizes the distribution of the latest supported OpenWrt releases and their corresponding Linux kernel versions across these 100 devices.

\begin{table}[]
\centering
\scriptsize
\caption{Distribution of the latest supported OpenWrt versions and associated Linux kernel versions for the 100 device–SoC pairs with available community support information. (*Recently dropped support)}
\begin{tabular}{|rr|r|}
\hline
\rowcolor[HTML]{C0C0C0} 
\multicolumn{1}{|r|}{\cellcolor[HTML]{C0C0C0}\textbf{\begin{tabular}[c]{@{}r@{}}Latest supported \\ OpenWrt version\end{tabular}}} & \textbf{\begin{tabular}[c]{@{}r@{}}Linux kernel \\ version associated\end{tabular}} & \textbf{\# devices} \\ \hline
\multicolumn{1}{|r|}{24.10.5} & 6.6.119 & 17 \\ \hline
\multicolumn{1}{|r|}{24.10.4} & 6.6.110 & 48 \\ \hline
\multicolumn{1}{|r|}{24.10.3} & 6.6.104 & 8 \\ \hline
\multicolumn{1}{|r|}{24.10.2} & 6.6.93 & 1 \\ \hline
\rowcolor[HTML]{FFFFFF} 
\multicolumn{1}{|r|}{\cellcolor[HTML]{FFFFFF}23.05*} & 5.15 & 10 \\ \hline
\rowcolor[HTML]{FFFFFF} 
\multicolumn{1}{|r|}{\cellcolor[HTML]{FFFFFF}19.07*} & 4.14 & 16 \\ \hline
\multicolumn{2}{|r|}{\textbf{Total}} & \multicolumn{1}{r|}{\textbf{100}} \\ \hline
\end{tabular}

\label{openwrt-support-distribution}
\end{table}

\subsection{Collecting regulation data}

We examine widely adopted compliance frameworks for consumer IoT and SOHO devices, focusing on ETSI EN 303 645~\cite{etsi303645} and NIST IR 8259~\cite{nistir8259}, which articulate baseline security expectations around vulnerability handling, software updates, and transparency. While many countries have introduced national standards and certification schemes for consumer IoT and SOHO devices, recent survey reports show that these efforts largely derive their security requirements from ETSI EN 303 645 and NIST IR 8259, adapting them to local regulatory contexts rather than introducing fundamentally new technical constraints~\cite{csa-iot-standards-2025}. As a result, these two frameworks effectively serve as the common foundation for compliance-oriented security governance across jurisdictions. Thus, we analyze their specifications and guidance documents to understand how current regulatory approaches address software maintenance, vulnerability remediation, and lifecycle support in deployed devices, and to assess the scope and enforcement boundaries of compliance-based governance.

Finally, we use this data to know if individual user action, community effort, or regulatory compliance potentially mitigates the kernel lock-in issue of the SOHO supply chain.

%% file: limitations.tex
A common issue in SOHO ecosystems is that SOHO device vendors reuse the same device model name across multiple hardware revisions, where different revisions may change the underlying SoC, SDK lineage, firmware instruction, and GPL releases. We therefore treat hardware revisions as first-class identifiers when they are available. We link a firmware image, GPL source release, SoC, and SDK to the same device only when their revision indicators are present and consistent across artifacts.

We emphasize that our firmware-source-based CVE estimation provides tight, device-specific \emph{estimates} of which known kernel vulnerabilities are likely present in the released kernel source tree, but it does not directly establish exploitability of these CVEs in deployed binaries. In particular, when we detect a CVE as ``present'' in the device firmware, we mean that the corresponding vulnerable code pattern appears to persist in the vendor-released firmware source tree and that the mainstream upstream kernel fix (as captured by our templates) does not appear to be applied in that source. Additionally, our current patch inference is scoped to mainstream upstream fixes, and we do not detect vendor-specific custom patches that may mitigate a vulnerability without matching the upstream fix pattern. Although we cannot systematically detect non-mainline, vendor-specific custom patches, we expect such silent mitigations to be relatively uncommon in practice.

We note that we might have missed some security- and privacy-related Reddit discussions since they might not use specific words. However, our machine learning model was trained and applied at the comment level, using the full text of each comment. Thus we ensure that security and privacy-related comments are robustly classified even when such concerns are not explicitly understandable. Furthermore, some inaccuracy is associated with machine learning models used to identify potential security and privacy-related comments. We attempted to reduce this concern by hyperparameter adjustment, rigorous model selection, manual validation and checking for saturation while creating our themes via qualitative coding, in line with best practices. 

In summary, despite these limitations, our analysis reveals meaningful and potentially generalizable insights into how users engage with security and privacy risks in SOHO devices, grounded in an ecologically valid, large-scale dataset.

%% file: RQ/RQ4.tex
In this section, we examine how governance around SOHO device security is mediated across the ecosystem, spanning end users, community-driven firmware efforts, and formal regulatory frameworks.

\subsection{Mediation by user demand around SOHO device security and privacy}

Building on our identification of security and privacy related comments, we investigate how users engage with device-related security risks in practice. Table~\ref{four_level_hierarchy} (of Appendix~\ref{app_rq4_dc_themes}) summarizes the resulting 4-level  hierarchy from our analysis. In this section, we restrict our discussion to user engagement patterns related to security and privacy of SOHO devices.

\vspace{1mm}
\noindent
\textbf{Configuration issues}: The most prominent form of user engagement relates to configuration issues in SOHO devices. Users frequently discuss incorrect settings, multi-device setup challenges, and uncertainty around enabling or disabling specific security-related features. These discussions often involve detailed descriptions of existing configurations and requests for validation or correction from the community. For example, a user describing a complex multi-device setup explains how they had to disable access points on multiple cameras, reposition devices, and reconfigure network components to restore functionality, noting that “\emph{… All 12 cameras now stream. I had to move some around to effectively work around the age-related defects. -Disabled the APs on all 5 DSC-833Ls. Having same SSID as router they just cause interference becasue cameras are not smart enuf to roam. -Cameras with password issues were moved to exterior views -Moved the 2.4ghz Extender so it was no longer blocked by a STEEL DOOR! … Keeping only AgentDVR running has reduced the problem of views switching off/on. Now sometimes 4 or 5 will switch OFF and on simultaneously, but not every few seconds like before}.” Other discussions reflect concerns about incorrect settings, where users caution that misconfiguration itself may cause security issues--- “\emph{You may be changing the settings and effectively blocking yourself -- this would occur with any brand. I wouldn't blame Dlink. What are your current IP address settings for your routers? And for your computer (including DNS)? Note that you should only be changing the LAN settings and not the WAN (internet side) addresses. ...}.” Configuration-related engagement also includes detailed peer guidance on setting up multiple routers or access points correctly. For instance, a user explains how to align SSIDs and security modes across routers while avoiding channel overlap, advising that “\emph{... I understand what this means. It means give the 2nd router the same SSID (WiFi name), password and security mode as the primary router. However set the channels different.... If not then something is missconfigured. Also I see that the 2730U has a WAN/LAN port....}.” Collectively, these interactions illustrate how users rely heavily on community feedback to diagnose and correct configuration problems, particularly in environments involving multiple devices and constrained firmware capabilities.

\vspace{1mm}
\noindent
\textbf{Network security}: Users also engage with security issues at the network level, often focusing on firewall behavior and network monitoring. Users describe observable symptoms such as unexpected traffic or connectivity issues and seek help interpreting whether these indicate genuine security risks or benign behavior. For example, one user explains how network isolation can mitigate exposure in shared connectivity settings, stating that “\emph{I understand it as this: LANDLORD HUB, provides shared internet across a number of people/flats/whatever 1.1) - YOUR D-link DIR-822, which you've physically connected to the landlord HUB, 1.1.1) -- YOUR RPI, connected to the YOUR DIR-822 1.1.2) -- Your other devices you should never connect any devices to the landlord’s hub, apart from your D-link… all devices should connect to your own router… ideally on a separate subnet… this should make your devices ‘inside’ your own network invisible to devices other than the landlord hub and the internet…},” while emphasizing the need to verify IP assignments and DNS configuration. In other cases, users highlight firewall-induced isolation between network segments, noting that “\emph{…if you use the TP-Link’s WAN/Internet port, it’ll get Internet but it’s firewall will stop things on the D-Link’s network from accessing things on the TP-Link’s network…}” and describing how reconfiguring the device as an access point disables the firewall to restore internal connectivity.

\vspace{1mm}
\noindent
\textbf{Awareness}: Users frequently engage through awareness-driven discussions that focus on device capabilities, security features, and available alternatives. Such engagement often draws on personal experience, with users sharing working setups, recommending alternative firmware, or clarifying credential-related practices that may affect security and privacy. For example, a user advises bypassing the vendor firmware entirely, stating that “\emph{Don't use the Dlink firmware to load the OpenWrt image. You probably need to bypass the dlink firmware with a recovery mode, and load the image that way},” while another highlights credential management practices by noting that “\emph{DLink routers have 3 different passwords. for the Router Admin. for Users to connect thru (this is the one you use on client machines). for Guest Access. Hard reset and reconfigure from scratch.}.”

\vspace{1mm}
\noindent
\textbf{Exploit}: Discussion of exploitation-related issues is uncommon. Only a small subset of discussions focus on how vulnerabilities may be exploited in practice or describe concrete attack scenarios. In these cases, users sometimes directly reference known weaknesses or attack surfaces, for example warning that “\emph{…these devices have a lot of vulns and should not be used in any other mode than bridge},” or noting that “\emph{…it looks like you have an HNAP capable router… so it’s exploitable regardless of your strong password}.” Other users contextualize exploitability by emphasizing attacker locality, explaining that “\emph{…the vulnerability comes down to attacker must be local to the network, so no remote vulnerabilities}.”

\vspace{1mm}
\noindent
\textbf{Update cycle}: Similarly, we observe relatively limited user engagement around update-cycle concerns, such as people being concerned about firmware update frequency, missing patches, or device EoL status. When updates are discussed, they often appear as secondary troubleshooting steps rather than as part of a sustained discussion on long-term security maintenance. For example, a user frames firmware freshness as a routine check and downplays potential risk by reasoning that “\emph{…the router is only 5 years old… WPA2 is much older than that… you should see if there’s a newer firmware from their support website},” while attributing observed problems to DNS rather than outdated software. In contrast, a smaller subset of users explicitly highlights patch neglect and its consequences, expressing frustration that “\emph{…they quickly ignore updates on less exciting models and then I’m stuck with vulnerabilities or looking for opensource firmware},” and describing how they repurpose unsupported devices as access points or abandon them entirely. Some discussions provide concrete temporal context to outdated support, noting that “\emph{…the firmware was last updated in February of 2013… there have been a few security issues with home routers since then that firmware updates have mitigated},” thereby implicitly linking missing patches to unresolved vulnerabilities. Other users compare vendors based on perceived update cadence, frequently inferring stronger long-term security support from higher-priced devices; for example, a user notes that “\emph{Ubiquiti and Linksys have regular security updates}” when recommending some other costlier alternatives. 

\vspace{1mm}

\noindent
\textbf{Summary}: Overall, these discussions suggest that update-cycle considerations are neither consistently understood nor central to user engagement. Many users infer security from device age, treat updates as incidental configuration steps, or lack clear expectations about vendor patch support, while only a minority explicitly reason about missing patches, end-of-life status, or long-term vulnerability exposure.

\subsection{Mediation by community-driven efforts}

Community-driven firmware projects such as OpenWrt partially mediate governance gaps left by vendors by offering alternate firmware, thereby extending device lifetimes and maintaining newer kernel baselines. However, this mediation is limited in both coverage and effectiveness. In our dataset of 306 devices, fewer than half have identifiable OpenWrt support, and only 100 devices can be mapped to the latest supported OpenWrt release, suggesting that community support is available for a subset of devices rather than uniformly across the ecosystem.
The dependence on proprietary vendor drivers and SDK components is a significant obstacle  to community mediation, as it limits the community support for newer kernels and the full functionality of the device. 

Thus, community efforts can only provide security improvements and support for a subset of devices that provide access to their drivers or use community-driven SDK baselines, such as is the case in Qualcomm, where their QSDK is derived from OpenWrt and is observed to do much better in terms of the supported Linux kernel~\cite{codelinaro2026qsdk}. They cannot fully fix upstream design choices that lock devices to old, hard to manage software bases but a good point to start with.

\subsection{Mediation by current regulations and compliance framework}

Our findings indicate that, while existing compliance frameworks are directionally aligned with improving SOHO device security, they do not meaningfully constrain the supply-chain dynamics that lead to kernel lock-in and long term vulnerability inheritance. In particular, compliance thresholds can be satisfied even when devices ship with Linux kernel baselines that are already end-of-life. 
Neither ETSI EN 303 645 \cite{etsi303645} nor NIST IR 8259 \cite{nistir8259} explicitly constrains the kernel baseline choices at the SoC-vendor, ODM/OEM, or device-vendor stages of the supply chain, nor do they explicitly require verifiable evidence that upstream support exists for the core operating system shipped on the device. As a result, compliance may not meaningfully address the upstream anchoring effects that we observe, where kernel choices made early in the supply chain propagate downstream and limit the feasibility of sustained security maintenance. Our observation is consistent with insights from related ecosystems like mobile security, where update delivery and security failures can depend on multiple actors, including third-party dependencies---not only on the final device vendor~\cite{ftc_mobile_security_updates,ftc_blu}.

From the user perspective, compliance signals are largely opaque. End users cannot realistically verify kernel baselines, SDK lineage, patch propagation. Compliance labels, where present, do not expose whether a device is built on a maintainable software foundation. To that end, adjacent efforts such as consumer-protection enforcement and the FCC's U.S.\ Cyber Trust Mark might strengthen device-security incentives and improve market-facing signals. However, even the U.S.\ Cyber Trust Mark is voluntary and does not by itself assign verifiable kernel-maintenance responsibility across SoC vendors, ODM/OEMs, and device vendors~\cite{us_cyber_trust_mark}.

Hopefully emerging measures such as the EU Cyber Resilience Act and the UK Cyber Security and Resilience Bill will further strengthen cybersecurity governance. However, at this point, it might be best to avoid speculating on their practical impact on SOHO kernel lock-in and wait as a community to observe their effects in practice~\cite{eu_cra,uk_csr_bill}.
This is because, previous work demonstrated that practical impact of regulation depends heavily on how requirements are operationalized and enforced~\cite{utz2019uninformed}

In summary, our analysis suggests that while existing compliance frameworks establish important high-level principles, stricter and more testable constraints are needed for compliance to meaningfully protect users in SOHO ecosystems.

%% file: implications.tex
Our analysis of SOHO devices reveals several important implications for how security responsibility and accountability are understood in consumer embedded systems. %

\vspace{1mm}
\noindent
\textbf{Vulnerability exposure is shaped upstream, not at the device alone:} Our findings show that vulnerability exposure in deployed SOHO devices is largely determined by upstream platform choices, particularly Linux kernel baselines inherited in SoC vendor developed SDKs. Once a kernel baseline is selected at the SDK stage, it is typically propagated through ODM/OEM integration and into shipped firmware without further upgrades. As a result, vulnerability debt is often introduced even before device-specific customization occurs and persists downstream across multiple products and vendors. Thus, treating security outcomes as a device vendor-only responsibility overlooks the constraints being imposed earlier in the supply chain.

\vspace{1mm}
\noindent
\textbf{Kernel version–centric security signals are insufficient:} We find that commonly used indicators of device security—such as kernel version strings or aggregate CVE counts—can misrepresent the true security posture of SOHO devices. Version-only analysis fails to capture backported fixes, configuration-dependent CVE applicability, and if a kernel baseline is still upstream-supported. More importantly, it obscures whether a device is built on a software foundation that can plausibly receive security updates over its lifetime. 

\vspace{1mm}
\noindent
\textbf{Kernel upgrades are not treated as a default maintenance path:} Our results show that kernel upgrades are rare in practice, with most devices remaining on inherited SDK baselines across updates. Even when devices continue to receive firmware releases, these updates typically do not involve detaching from older kernel tracks. This suggests that kernel modernization is not integrated into routine maintenance workflows, reinforcing long-term exposure when SDK baselines are outdated or end-of-life.

\vspace{1mm}
\noindent
\textbf{Users have limited ability to assess or influence device security:} From the user  perspective, our analysis of Reddit discussions shows that only a small fraction of users raise security and privacy concerns, and these discussions are mostly limited to configuration issues rather than long-term risks such as vulnerability patching or end-of-life support. Users lack visibility into kernel baselines, SDK lineage, and vulnerabilities, making it difficult to evaluate device security or exert market pressure. 

\vspace{1mm}
\noindent
\textbf{Sustained kernel maintenance is feasible under alternate models:} The contrast between SDK anchored firmware and community derived baselines (such as OpenWrt) demonstrates that sustained kernel maintenance is technically feasible on comparable hardware when platform and driver maintenance models permit it. This indicates that persistent use of older kernels is not inevitable and that plausible upgrade paths exist, but they depend on a combination of upstream choices and downstream integration incentives.

%% file: recommendation.tex
Based on these implications, we outline actionable recommendations for key stakeholders of the SOHO device ecosystem. %

\vspace{1mm}
\noindent
\textbf{Recommendations for compliance bodies and regulators:} Existing compliance frameworks, such as ETSI EN 303 645 and NIST IR 8259, establish important high-level principles but do not directly constrain kernel baseline selection or upstream supportability. We recommend that compliance requirements explicitly mandate that shipped firmware use a Linux kernel baseline that is upstream-supported at the time of ship, with preference for LTS kernels when available. If ``end-of-life'' declared components are still being used by a stakeholder, compliance should require an explicit maintenance undertaking that mandates that security patches will be sustained over the device’s declared support window by the stakeholder. That is, if an ODM/OEM uses an SDK with an end-of-life Linux kernel, then the ODM/OEM (or the device vendor that subcontracts them) will explicitly assume responsibility for maintaining and patching the kernel baseline even after the mainstream Linux kernel has dropped support for it. Such constraints would make compliance more effective in addressing supply chain-driven vulnerability inheritance.

Furthermore, Software Bill of Materials (SBOM) requirements should be paired with an accompanying security status report rather than relying on raw component listings alone. We recommend that devices provide a concise, user-facing checklist indicating whether components like  Linux kernel are supported and actively maintained. For auditors and technical stakeholders, the same report can include more detailed information on kernel baselines, support windows, and patch posture, enabling verification without requiring users to interpret GPL source trees or kernel build artifacts.

\vspace{1mm}
\noindent
\textbf{Recommendation for the use of community alternatives:}
Based on our analysis, we find that devices that derive their SDK kernel baseline from community alternatives such as OpenWrt have ``better'' kernels in their shipped firmware with more visibility to the end user and longer community-driven support. This adoption has been observed in the SoC vendor stage, but can also be integrated at the ODM/OEM stage.

\vspace{1mm}
\noindent
\textbf{Recommendations for ODMs and device vendors:} ODM and device vendors should treat kernel baseline selection as an explicit integration and release gate rather than inheriting SDK defaults without scrutiny. “Shipping what the SDK provides” should not be considered sufficient for security assurance. Vendors should require evidence that the inherited kernel baseline can be maintained throughout the intended life cycle and should prioritise platforms that align with supported kernel tracks. In case of a lack of this evidence from the SoC vendor's SDK, the ODMs can choose to upgrade the kernel baseline to a ``better'' and well-supported kernel.

\vspace{1mm}
\noindent
\textbf{Recommendations for SoC vendors:} Although SoC vendors may not be directly regulated by device-level compliance regimes, their SDK choices have a disproportionate downstream impact. We recommend that SoC vendors align SDK baselines with upstream-supported and preferably LTS kernel lines. Doing so reduces friction for downstream compliance, enables sustained security maintenance, and lowers vulnerability inheritance across multiple vendors and products. If one SoC vendor chooses to make a ``better'' choice for even the SoC SDK the impact is cascading and improves security for end users across device vendors.%

While our empirical measurements focus on the Linux kernel, these recommendations generalise to other widely reused third-party components in the SOHO firmware stack. The kernel serves as a concrete and measurable anchor for demonstrating how vulnerability debt is introduced and inherited, motivating broader compliance and accountability improvements across the ecosystem.

%% file: conclusion.tex
This study examined the security posture of SOHO devices through an end-to-end supply-chain perspective, combining firmware analysis, vendor GPL source inspection, and user-centric evidence from public online discussions.We correlate shipped firmware with GPL releases and SDK lineage. Our analysis shows that vulnerability exposure in deployed devices is shaped by upstream constraints, especially long-lived Linux kernel baselines inherited from SoC SDKs, rather than by isolated vendor choices. As a result, vulnerability debt is often introduced early in the supply chain and persists downstream to end users. Our analysis further demonstrates that version-centric vulnerability estimates can substantially misrepresent real-world device security. By incorporating GPL-based source analysis and build evidence, we provide a more precise view of which kernel vulnerabilities are actually present and whether fixes are incorporated. At the same time, kernel upgrades are rare, and most devices stay on the same SDK-derived kernel across updates. The contrast with OpenWrt-derived baselines highlights that sustained kernel maintenance is feasible under alternative platform and driver maintenance models. From the user perspective, our large-scale analysis of Reddit discussions shows that only a small fraction of users raise security and privacy concerns. They are often limited to configuration issues rather than long-term risks such as patching and end-of-life support. This suggests that users have limited ability to assess or influence device security outcomes on their own.

Overall, our findings point to a misalignment between where vulnerability debt is introduced in the SOHO ecosystem and where responsibility is typically assumed to lie. Addressing this gap will require coordinated efforts across the supply chain, including greater transparency around SDK baselines and stronger incentives for sustained kernel maintenance. We hope this work contributes empirical grounding to ongoing discussions on accountability and security in consumer embedded systems.

%% file: ethical_consideration.tex
We considered whether formal ethical review was required for our study protocol. We discussed our protocol with our Institute Ethics Committee (IEC). They indicated that ethical approval was not required, as our study exclusively collected and analyzed data from publicly accessible sources, including vendor-published firmware images, GPL source code releases, and publicly available online discussions. However, we recognize that this discussion, by itself, does not guarantee ethical research. The collection and analysis of public data can still raise ethical concerns, particularly when such data was not created with research purposes in mind~\cite{BUCK2021102655}. To this end, we made every effort to conduct our research responsibly, seeking to minimize potential harm and maximize potential benefits for relevant stakeholders, including device vendors, SoC vendors, end users, and the platforms and ecosystems of these devices.

In line with prior work on firmware analysis, we relied exclusively on publicly available data from their respective owners/forums. Our technical analysis was restricted to firmware images and GPL source releases published by vendors through official support portals, as well as publicly documented metadata associated with these releases. We strictly adhered to the ethical principles outlined in previous studies~\cite{Eysenbach1103}, avoiding any private communications or data collection behind authentication walls. We did not collect any private data, bypass authentication mechanisms, or interact with deployed devices on live networks. All analysis was performed offline without exploiting any deployed systems.

Although the use of publicly available firmware and source code is common in security research~\cite{BUCK2021102655}, ethical concerns can still arise, particularly when research findings may expose systemic weaknesses or affect stakeholders who did not explicitly consent to participation. To mitigate these risks, we took several precautions. We did not focus our analysis on individual end users or specific device deployments, nor did we attempt to attribute responsibility to particular engineers or development teams. We try to detect known and publicly disclosed vulnerabilities in these devices to determine where the problem arises, and do not try to find or disclose any unknown vulnerabilities that may exist on these devices, which could lead to exploitation attempts. Our study does not discover or disclose any new vulnerabilities or CVEs. Rather, we use already-public CVE records to assess how known, patched vulnerabilities may persist across vendor-released firmware source artifacts. Every CVE considered in our analysis has already been publicly documented through the CVE/NVD ecosystem and associated with upstream mitigation or patch information. Instead, our analysis emphasizes ecosystem-level patterns, such as inherited kernel baselines, SDK constraints, and supply-chain dependencies, rather than isolated implementation flaws. We also avoid publicly releasing fine-grained information that directly maps individual CVEs to specific device models in a way that could unnecessarily increase risk. Instead, we present aggregate and ecosystem-level results in the paper, while using detailed mappings only for responsible notification to relevant stakeholders.

In order to ensure responsible notification, we contacted all affected device vendors through their dedicated vulnerability reporting or security contact channels. For each device vendor, we described who we are, the purpose of our study, and our methodology, and privately provided a detailed list of the device models we analyzed together with the known CVEs that our pipeline indicated may be present in the corresponding released source artifacts. Where available, we also provided pointers to general upstream patches or mitigation information for the relevant CVEs. Three out of the five contacted device vendors responded. Their responses acknowledged the report and indicated that End-of-Life (EoL) or End-of-Service (EoS) devices may no longer receive firmware updates, while also noting that other reported cases were being investigated internally.

Since our results suggest that vulnerability debt may originate upstream through SoC SDK kernel choices, we also contacted the relevant SoC vendors. These notifications did not claim that the SoC vendor was solely responsible for any downstream product vulnerability. Rather, we informed them that products using SoCs associated with their SDK lineage appeared to contain known CVEs, and we provided the corresponding list of CVEs observed in products using those SoC/SDK lineages. Our goal was to support internal review, SDK maintenance, downstream vendor guidance, and clarification of patch responsibility across the supply chain.

Our study also includes an analysis of user discussions drawn from Reddit, a public online forum where content is openly accessible without authentication. These discussions reflect unsolicited, naturally occurring user experiences and perceptions of SOHO routers and IP camera devices. While users do not explicitly consent to having their posts analyzed for research, such discussions are publicly visible and have been widely studied in prior work~\cite{nguyen2019short,raj2024just,akgul2024decade}. To reduce potential harm, we analyzed comments only in aggregate, avoided targeting specific individuals or communities, and used anonymized excerpts solely for illustrative purposes. We do not associate quotes to any usernames or personal identifiers. Our analysis focuses on emergent themes rather than individual users or posts. We did not attempt to infer user identities, locations, or personal characteristics beyond what was explicitly stated in the public text.

All collected data was stored securely with access limited to the research team. We minimized personal identifier collection and retained only data needed for our research questions. In order to minimize risks or harms to deployed systems as well as end users, we used public artifacts, avoided interaction with deployed systems, reported aggregate ecosystem-level results, and privately notified relevant stakeholders where appropriate. We believe that our findings contribute constructive insights that can support more informed discussions among vendors, SoC vendors, platform providers, policymakers, and users.

In summary, we strongly believe that our research was conducted ethically and responsibly.  We hope that the insights presented in this work can help improve security practices, inform realistic expectations around device maintenance and lifecycle support, and ultimately benefit the broader SOHO device ecosystem.

%% file: openscience.tex
Our work adheres to the open science policy. We are releasing the artifacts (models) with accompanying code so that the results of this work can be reproduced and our models can be used in further research (e.g., for discovering security and privacy-related comments from discusions around SOHO devices from Reddit). Since the raw data (Reddit and other forum datasets) might contain identifying information related to security and privacy concerns about various SOHO devices, to address potential ethical concerns about open science compliance,
we decided to share the raw datasets with verified researchers upon request, rather than releasing them publicly. The instructions on how to request this data is present in the Zenodo link below. In addition, as part of this work, we also release all relevant firmware-related data, together our scripts to help uncover the entities of supply chain for soho device firmwares
at the following link: 
\url{https://doi.org/10.5281/zenodo.20433799}

%% file: references.bib
@misc{fcc_equip_auth,
  title        = {Equipment Authorization},
  author       = {{Federal Communications Commission}},
  howpublished = {\url{https://www.fcc.gov/engineering-technology/laboratory-division/general/equipment-authorization}},
  year         = {2025},
  note         = {Accessed: 2026-02-05},
}

@misc{binwalkGitHub,
  title        = {{Binwalk: Firmware Analysis Tool}},
  author       = {{ReFirmLabs}},
  year         = {2025},
  note         = {Accessed: Dec. 7, 2025},
  howpublished = {\url{https://github.com/ReFirmLabs/binwalk}}
}

@misc{openwrt_toh2026,
  title        = {OpenWrt Table of Hardware (TOH)},
  author       = {{OpenWrt Project}},
  howpublished = {\url{https://toh.openwrt.org/?view=normal}},
  year         = {2026},
  note         = {Accessed: 2026-02-02},
  url          = {https://toh.openwrt.org/?view=normal},
}

@misc{embaGitHub,
  title        = {{EMBA: The firmware security analyzer}},
  author       = {{e-m-b-a}},
  year         = {2025},
  url          = {https://github.com/e-m-b-a/emba},
  note         = {Accessed: Dec. 7, 2025},
  howpublished = {\url{https://github.com/e-m-b-a/emba}}
}

@misc{unblobOrg,
  title        = {{unblob: Accurate and fast extraction suite for binary blobs}},
  author       = {{ONEKEY}},
  year         = {2025},
  note         = {Accessed: Dec. 7, 2025},
  url          = {https://unblob.org/},
  howpublished = {\url{https://unblob.org/}}
}

@InProceedings{10.1145/3617072.3617110,
author = {Chalhoub, George and Martin, Andrew},
title = {But is it exploitable? Exploring how Router Vendors Manage and Patch Security Vulnerabilities in Consumer-Grade Routers},
year = {2023},
isbn = {9798400708145},
publisher = {Association for Computing Machinery},
address = {New York, NY, USA},
url = {https://doi.org/10.1145/3617072.3617110},
doi = {10.1145/3617072.3617110},
booktitle = {Proceedings of the 2023 European Symposium on Usable Security},
pages = {277–295},
numpages = {19},
location = {Copenhagen, Denmark},
series = {EuroUSEC '23}
}

@misc{dlinkWebsite,
  title        = {{D-Link: Smart Home, SMB and Enterprise Networking Solutions}},
  author       = {{D-Link Systems, Inc.}},
  year         = {2026},
  note         = {Accessed: Jan. 26, 2026},
  howpublished = {\url{https://www.dlink.com/in/en}}
}

@misc{netgearWebsite,
  title        = {{NETGEAR: Advanced WiFi \& Networking Products}},
  author       = {{Netgear, Inc.}},
  year         = {2026},
  note         = {Accessed: Jan. 26, 2026},
  howpublished = {\url{https://www.netgear.com/}}
}

@misc{tplinkWebsite,
  title        = {{TP-Link: Networking Products and Solutions}},
  author       = {{TP-Link Technologies Co., Ltd.}},
  year         = {2026},
  note         = {Accessed: Jan. 26, 2026},
  howpublished = {\url{https://www.tp-link.com/in/}}
}

@misc{linksysWebsite,
  title        = {{Linksys: Home and Business Wi-Fi and Networking}},
  author       = {{Linksys, Inc.}},
  year         = {2026},
  note         = {Accessed: Jan. 26, 2026},
  howpublished = {\url{https://www.linksys.com/}}
}

@misc{trendnetWebsite,
  title        = {{TRENDnet: Networking and Surveillance Solutions}},
  author       = {{TRENDnet, Inc.}},
  year         = {2026},
  note         = {Accessed: Jan. 26, 2026},
  howpublished = {\url{https://www.trendnet.com/home}}
}

@article{chung2024scaling,
  title={Scaling instruction-finetuned language models},
  author={Chung, Hyung Won and Hou, Le and Longpre, Shayne and Zoph, Barret and Tay, Yi and Fedus, William and Li, Yunxuan and Wang, Xuezhi and Dehghani, Mostafa and Brahma, Siddhartha and others},
  journal={Journal of Machine Learning Research},
  volume={25},
  number={70},
  pages={1--53},
  year={2024}
}

@InProceedings{10.1007/978-3-031-35504-2_10,
author="Helmke, R.
and vom Dorp, J.",
editor="Gruss, Daniel
and Maggi, Federico
and Fischer, Mathias
and Carminati, Michele",
title="Extended Abstract: Towards Reliable and Scalable Linux Kernel CVE Attribution in Automated Static Firmware Analyses",
booktitle="Detection of Intrusions and Malware, and Vulnerability Assessment",
year="2023",
publisher="Springer Nature Switzerland",
address="Cham",
pages="201--210",
isbn="978-3-031-35504-2"
}

@inproceedings{10.1145/3533767.3534366,
author = {Zhao, Binbin and Ji, Shouling and Xu, Jiacheng and Tian, Yuan and Wei, Qiuyang and Wang, Qinying and Lyu, Chenyang and Zhang, Xuhong and Lin, Changting and Wu, Jingzheng and Beyah, Raheem},
title = {A large-scale empirical analysis of the vulnerabilities introduced by third-party components in IoT firmware},
year = {2022},
isbn = {9781450393799},
publisher = {Association for Computing Machinery},
address = {New York, NY, USA},
url = {https://doi.org/10.1145/3533767.3534366},
doi = {10.1145/3533767.3534366},
booktitle = {Proceedings of the 31st ACM SIGSOFT International Symposium on Software Testing and Analysis},
pages = {442–454},
numpages = {13},
location = {Virtual, South Korea},
series = {ISSTA 2022}
}

@article {Eysenbach1103,
	author = {Eysenbach, Gunther and Till, James E},
	title = {Ethical issues in qualitative research on internet communities},
	volume = {323},
	number = {7321},
	pages = {1103--1105},
	year = {2001},
	doi = {10.1136/bmj.323.7321.1103},
	publisher = {BMJ Publishing Group Ltd},
	issn = {0959-8138},
	URL = {https://www.bmj.com/content/323/7321/1103},
	eprint = {https://www.bmj.com/content/323/7321/1103.full.pdf},
	journal = {BMJ}
}

@article{BUCK2021102655,
title = {I didn't sign up for your research study: The ethics of using “public” data},
journal = {Computers and Composition},
volume = {61},
pages = {102655},
year = {2021},
note = {Rhetorics of Data: Collection, Consent, \& Critical Digital Literacies},
issn = {8755-4615},
doi = {https://doi.org/10.1016/j.compcom.2021.102655},
url = {https://www.sciencedirect.com/science/article/pii/S8755461521000323},
author = {Amber M. Buck and Devon F. Ralston}
}

@inproceedings{10.5555/3620237.3620429,
author = {Zhao, Binbin and Ji, Shouling and Zhang, Xuhong and Tian, Yuan and Wang, Qinying and Pu, Yuwen and Lyu, Chenyang and Beyah, Raheem},
title = {UVSCAN: detecting third-party component usage violations in IoT firmware},
year = {2023},
isbn = {978-1-939133-37-3},
publisher = {USENIX Association},
address = {USA},
booktitle = {Proceedings of the 32nd USENIX Conference on Security Symposium},
articleno = {192},
numpages = {18},
location = {Anaheim, CA, USA},
series = {SEC '23}
}

@techreport{weidenbach-2020,
	author = {Weidenbach, Peter and Dorp, Johanne vom},
	month = {6},
    institution = {FRAUNHOFER-INSTITUT FÜR KOMMUNIKATION, INFORMATIONSVERARBEITUNG UND ERGONOMIE, FKIE},
	title = {{Home Router Security Report 2020}},
	year = {2020},
    
    note         = {Accessed: Jan. 26, 2026},
    howpublished = {\url{https://www.fkie.fraunhofer.de/content/dam/fkie/de/documents/HomeRouter/HomeRouterSecurity_2020_Bericht.pdf}},
   
}

@inproceedings{10.1145/3427228.3427294,
author = {Kim, Mingeun and Kim, Dongkwan and Kim, Eunsoo and Kim, Suryeon and Jang, Yeongjin and Kim, Yongdae},
title = {FirmAE: Towards Large-Scale Emulation of IoT Firmware for Dynamic Analysis},
year = {2020},
isbn = {9781450388580},
publisher = {Association for Computing Machinery},
address = {New York, NY, USA},
url = {https://doi.org/10.1145/3427228.3427294},
doi = {10.1145/3427228.3427294},
booktitle = {Proceedings of the 36th Annual Computer Security Applications Conference},
pages = {733–745},
numpages = {13},
location = {Austin, USA},
series = {ACSAC '20}
}

@inproceedings{raj2024just,
  author={Rohit Raj and Mridul Newar and Mainack Mondal},
  title={"I just hated it and I want my money back": Data-driven Understanding of Mobile VPN Service Switching Preferences in The Wild},
  year={2024},
  cdate={1704067200000},
  url={https://www.usenix.org/conference/usenixsecurity24/presentation/raj},
  booktitle={USENIX Security Symposium}
}

@inproceedings{ghosh2025wasn,
  author       = {Rajdeep Ghosh and
                  Shiladitya De and
                  Mainack Mondal},
  editor       = {Lujo Bauer and
                  Giancarlo Pellegrino},
  title        = {"I wasn't sure if this is indeed a security risk": Data-driven Understanding
                  of Security Issue Reporting in GitHub Repositories of Open Source
                  npm Packages},
  booktitle    = {34th {USENIX} Security Symposium, {USENIX} Security 2025, Seattle,
                  WA, USA, August 13-15, 2025},
  pages        = {2145--2164},
  publisher    = {{USENIX} Association},
  year         = {2025},
  url          = {https://www.usenix.org/conference/usenixsecurity25/presentation/ghosh},
  bibsource    = {dblp computer science bibliography, https://dblp.org}
}

@inproceedings{10.5555/3620237.3620518,
author = {Angelakopoulos, Ioannis and Stringhini, Gianluca and Egele, Manuel},
title = {FirmSolo: enabling dynamic analysis of binary Linux-based IoT kernel modules},
year = {2023},
isbn = {978-1-939133-37-3},
publisher = {USENIX Association},
address = {USA},
booktitle = {Proceedings of the 32nd USENIX Conference on Security Symposium},
articleno = {281},
numpages = {18},
location = {Anaheim, CA, USA},
series = {SEC '23}
}

@inproceedings{ChenWBE16,
  title = {Towards Automated Dynamic Analysis for Linux-based Embedded Firmware},
  author = {Daming D. Chen and Maverick Woo and David Brumley and Manuel Egele},
  year = {2016},
  url = {http://www.internetsociety.org/sites/default/files/blogs-media/towards-automated-dynamic-analysis-linux-based-embedded-firmware.pdf},
  researchr = {https://researchr.org/publication/ChenWBE16},
  cites = {0},
  citedby = {0},
  booktitle = {23nd Annual Network and Distributed System Security Symposium, NDSS 2016, San Diego, California, USA, February 21-24, 2016},
  publisher = {The Internet Society},
}

@Article{s23229221,
AUTHOR = {Zhou, Xu and Wang, Pengfei and Zhou, Lei and Xun, Peng and Lu, Kai},
TITLE = {A Survey of the Security Analysis of Embedded Devices},
JOURNAL = {Sensors},
VOLUME = {23},
YEAR = {2023},
NUMBER = {22},
ARTICLE-NUMBER = {9221},
URL = {https://www.mdpi.com/1424-8220/23/22/9221},
PubMedID = {38005606},
ISSN = {1424-8220},
DOI = {10.3390/s23229221}
}

@online{talos_vpnfilter_2018,
  author       = {W. Largent},
  title        = {New {VPNFilter} malware targets at least 500K networking devices worldwide},
  organization = {Cisco Talos Intelligence Group},
  year         = {2018},
  month        = may,
  day          = {23},
  url          = {https://blog.talosintelligence.com/vpnfilter/},
  note         = {Accessed: Dec. 7, 2025}
}

@online{doj_court_operation_botnet_2024,
  author       = {{U.S. Department of Justice, Office of Public Affairs}},
  title        = {Court-Authorized Operation Disrupts Worldwide Botnet Used by People{\textquoteright}s Republic of China State-Sponsored Hackers},
  organization = {U.S. Department of Justice},
  year         = {2024},
  month        = sep,
  day          = {18},
  url          = {https://www.justice.gov/archives/opa/pr/court-authorized-operation-disrupts-worldwide-botnet-used-peoples-republic-china-state},
  howpublished= {\url{https://www.justice.gov/archives/opa/pr/court-authorized-operation-disrupts-worldwide-botnet-used-peoples-republic-china-state}},
  note         = {Accessed: Dec. 7, 2025}
}

@online{jsca_prc_linked_botnet_2024,
  author       = {{Federal Bureau of Investigation} and {Cyber National Mission Force} and {National Security Agency}},
  title        = {People{\textquoteright}s Republic of China-Linked Actors Compromise Routers and IoT Devices for Botnet Operations},
  organization = {FBI, CNMF, and NSA},
  year         = {2024},
  month        = sep,
  day          = {18},
  url          = {https://media.defense.gov/2024/Sep/18/2003547016/-1/-1/0/CSA-PRC-LINKED-ACTORS-BOTNET.PDF},
  note         = {Joint Cybersecurity Advisory JCSA-20240918-001, Accessed: Dec. 7, 2025},
  howpublished ={\url{https://media.defense.gov/2024/Sep/18/2003547016/-1/-1/0/CSA-PRC-LINKED-ACTORS-BOTNET.PDF}}
  
}

@online{securityscorecard_volt_typhoon_2024,
  author       = {{SecurityScorecard STRIKE Team}},
  title        = {Threat Intelligence Research: Volt Typhoon Compromises 30\% of Cisco RV320/325 Devices in 37 Days},
  organization = {SecurityScorecard},
  year         = {2024},
  month        = jan,
  day          = {11},
  url          = {https://securityscorecard.com/blog/threat-intelligence-research-volt-typhoon/},
  howpublished= {\url{https://securityscorecard.com/blog/threat-intelligence-research-volt-typhoon/}},
  note         = {Accessed: Dec. 7, 2025}
}

@inproceedings{10.5555/2671225.2671232,
author = {Costin, Andrei and Zaddach, Jonas and Francillon, Aur\'{e}lien and Balzarotti, Davide},
title = {A large-scale analysis of the security of embedded firmwares},
year = {2014},
isbn = {9781931971157},
publisher = {USENIX Association},
address = {USA},
booktitle = {Proceedings of the 23rd USENIX Conference on Security Symposium},
pages = {95–110},
numpages = {16},
location = {San Diego, CA},
series = {SEC'14}
}

@INPROCEEDINGS{9152796,
  author={Redini, Nilo and Machiry, Aravind and Wang, Ruoyu and Spensky, Chad and Continella, Andrea and Shoshitaishvili, Yan and Kruegel, Christopher and Vigna, Giovanni},
  booktitle={2020 IEEE Symposium on Security and Privacy (SP)}, 
  title={Karonte: Detecting Insecure Multi-binary Interactions in Embedded Firmware}, 
  year={2020},
  volume={},
  number={},
  pages={1544-1561},
  doi={10.1109/SP40000.2020.00036}}

@misc{gundecha2018selenium,
  title        = {Selenium},
  author       = {{Selenium Contributors}},
  year         = {2026},
  url          = {https://www.selenium.dev/},
  note         = {Accessed: January 7, 2026},
  organization = {Software Freedom Conservancy}
}

@inproceedings{nguyen2019short,
  title={Short text, large effect: Measuring the impact of user reviews on android app security \& privacy},
  author={Nguyen, Duc Cuong and Derr, Erik and Backes, Michael and Bugiel, Sven},
  booktitle={2019 IEEE symposium on Security and Privacy (SP)},
  pages={555--569},
  year={2019},
  organization={IEEE}
}

@inproceedings{li2023s,
  title={“It’s up to the Consumer to be Smart”: Understanding the Security and Privacy Attitudes of Smart Home Users on Reddit},
  author={Li, Jingjie and Sun, Kaiwen and Huff, Brittany Skye and Bierley, Anna Marie and Kim, Younghyun and Schaub, Florian and Fawaz, Kassem},
  booktitle={2023 IEEE Symposium on Security and Privacy (SP)},
  pages={2850--2866},
  year={2023},
  organization={IEEE}
}

@article{devlin2018bert,
  title={Bert: Pre-training of deep bidirectional transformers for language understanding},
  author={Devlin, Jacob},
  journal={arXiv preprint arXiv:1810.04805},
  year={2018}
}

@article{he2020deberta,
  title={Deberta: Decoding-enhanced bert with disentangled attention},
  author={He, Pengcheng and Liu, Xiaodong and Gao, Jianfeng and Chen, Weizhu},
  journal={arXiv preprint arXiv:2006.03654},
  year={2020}
}

@article{liu2019roberta,
  title={Roberta: A robustly optimized bert pretraining approach},
  author={Liu, Yinhan},
  journal={arXiv preprint arXiv:1907.11692},
  volume={364},
  year={2019}
}

@misc{akiba2019optunanextgenerationhyperparameteroptimization,
      title={Optuna: A Next-generation Hyperparameter Optimization Framework}, 
      author={Takuya Akiba and Shotaro Sano and Toshihiko Yanase and Takeru Ohta and Masanori Koyama},
      year={2019},
      eprint={1907.10902},
      archivePrefix={arXiv},
      primaryClass={cs.LG},
      url={https://arxiv.org/abs/1907.10902}, 
      howpublished={\url{https://arxiv.org/abs/1907.10902}}
}

@book{saldana2015coding,
  title={The Coding Manual for Qualitative Researchers},
  author={Saldana, J.},
  isbn={9781473943599},
  url={https://books.google.co.in/books?id=ZhxiCgAAQBAJ},
  year={2015},
  publisher={SAGE Publications}
}

@article{saunders2018saturation,
  title={Saturation in qualitative research: exploring its conceptualization and operationalization},
  author={Saunders, Benjamin and Sim, Julius and Kingstone, Tom and Baker, Shula and Waterfield, Jackie and Bartlam, Bernadette and Burroughs, Heather and Jinks, Clare},
  journal={Quality \& quantity},
  volume={52},
  pages={1893--1907},
  year={2018},
  publisher={Springer}
}

@inproceedings{10.1145/2702123.2702561,
author = {Harboe, Gunnar and Huang, Elaine M.},
title = {Real-World Affinity Diagramming Practices: Bridging the Paper-Digital Gap},
year = {2015},
isbn = {9781450331456},
publisher = {Association for Computing Machinery},
address = {New York, NY, USA},
url = {https://doi.org/10.1145/2702123.2702561},
doi = {10.1145/2702123.2702561},
booktitle = {Proceedings of the 33rd Annual ACM Conference on Human Factors in Computing Systems},
pages = {95–104},
numpages = {10},
location = {Seoul, Republic of Korea},
series = {CHI '15}
}

@inproceedings{akgul2024decade,
  title={A Decade of $\{$Privacy-Relevant$\}$ Android App Reviews: Large Scale Trends},
  author={Akgul, Omer and Peddinti, Sai Teja and Taft, Nina and Mazurek, Michelle L and Harkous, Hamza and Srivastava, Animesh and Seguin, Benoit},
  booktitle={33rd USENIX Security Symposium (USENIX Security 24)},
  pages={5089--5106},
  year={2024}
}

@article{mukherjee2020empirical,
  title={An empirical study on user reviews targeting mobile apps' security \& privacy},
  author={Mukherjee, Debjyoti and Ahmadi, Alireza and Pour, Maryam Vahdat and Reardon, Joel},
  journal={arXiv preprint arXiv:2010.06371},
  year={2020}
}

@inproceedings{antonakakis2017understanding,
  title={Understanding the mirai botnet},
  author={Antonakakis, Manos and April, Tim and Bailey, Michael and Bernhard, Matt and Bursztein, Elie and Cochran, Jaime and Durumeric, Zakir and Halderman, J Alex and Invernizzi, Luca and Kallitsis, Michalis and others},
  booktitle={26th USENIX security symposium (USENIX Security 17)},
  pages={1093--1110},
  year={2017}
}

@INPROCEEDINGS{fernandes2016security,
  author={Fernandes, Earlence and Jung, Jaeyeon and Prakash, Atul},
  booktitle={2016 IEEE Symposium on Security and Privacy (SP)}, 
  title={Security Analysis of Emerging Smart Home Applications}, 
  year={2016},
  volume={},
  number={},
  pages={636-654},
  doi={10.1109/SP.2016.44}}

@inproceedings{nino2024unveiling,
  title={Unveiling $\{$IoT$\}$ security in reality: A $\{$Firmware-Centric$\}$ journey},
  author={Nino, Nicolas and Lu, Ruibo and Zhou, Wei and Lee, Kyu Hyung and Zhao, Ziming and Guan, Le},
  booktitle={33rd USENIX Security Symposium (USENIX Security 24)},
  pages={5609--5626},
  year={2024}
}

@inproceedings{vetrivel2023examining,
  title={Examining consumer reviews to understand security and privacy issues in the market of smart home devices},
  author={Vetrivel, Swaathi and Van Harten, Veerle and Ga{\~n}{\'a}n, Carlos H and Van Eeten, Michel and Parkin, Simon},
  booktitle={32nd USENIX security symposium (USENIX security 23)},
  pages={1523--1540},
  year={2023}
}

@article{adam2024survey,
  title={A survey on security, privacy, trust, and architectural challenges in IoT systems},
  author={Adam, Mumin and Hammoudeh, Mohammad and Alrawashdeh, Rana and Alsulaimy, Basil},
  journal={IEEE Access},
  volume={12},
  pages={57128--57149},
  year={2024},
  publisher={IEEE}
}

@article{deep2022survey,
  title={A survey of security and privacy issues in the Internet of Things from the layered context},
  author={Deep, Samundra and Zheng, Xi and Jolfaei, Alireza and Yu, Dongjin and Ostovari, Pouya and Kashif Bashir, Ali},
  journal={Transactions on Emerging Telecommunications Technologies},
  volume={33},
  number={6},
  pages={e3935},
  year={2022},
  publisher={Wiley Online Library}
}

@techreport{Kumaraguru_Cranor:2005,
  address = {Pittsburgh, PA},
  author = {Kumaraguru, Ponnurangam and Cranor, Lorrie Faith},
  institution = {Institute for Software Research International, School of Computer
	Science, Carnegie Mellon University},
  month = {Dezember},
  number = {CMU-ISRI-5-138},
  title = {{P}rivacy {I}ndexes: {A} {S}urvey of {W}estin's {S}tudies},
  year = 2005
}

@inproceedings{10.1145/2858036.2858214,
author = {Dupree, Janna Lynn and Devries, Richard and Berry, Daniel M. and Lank, Edward},
title = {Privacy Personas: Clustering Users via Attitudes and Behaviors toward Security Practices},
year = {2016},
isbn = {9781450333627},
publisher = {Association for Computing Machinery},
address = {New York, NY, USA},
url = {https://doi.org/10.1145/2858036.2858214},
doi = {10.1145/2858036.2858214},
booktitle = {Proceedings of the 2016 CHI Conference on Human Factors in Computing Systems},
pages = {5228–5239},
numpages = {12},
location = {San Jose, California, USA},
series = {CHI '16}
}

@article{10.1016/j.tele.2017.04.013,
author = {Barth, Susanne and de Jong, Menno D.T.},
title = {The privacy paradox Investigating discrepancies between expressed privacy concerns and actual online behavior A systematic literature review},
year = {2017},
issue_date = {November 2017},
publisher = {Pergamon Press, Inc.},
address = {USA},
volume = {34},
number = {7},
issn = {0736-5853},
url = {https://doi.org/10.1016/j.tele.2017.04.013},
doi = {10.1016/j.tele.2017.04.013},
journal = {Telemat. Inf.},
month = nov,
pages = {1038–1058},
numpages = {21}
}

@techreport{csa-iot-standards-2025,
  title        = {Consumer IoT Device Cybersecurity Standards, Policies, and Certification Schemes 2025},
  author       = {{Connectivity Standards Alliance (CSA)}},
  institution  = {Connectivity Standards Alliance},
  year         = {2025},
  type         = {Technical Report},
  url          = {https://csa-iot.org/wp-content/uploads/2025/06/Consumer-IoT-Device-Cybersecurity-Standards-Policies-and-Certification-Schemes-2025-_FINAL.pdf},
  
}

@misc{nistir8259,
  author       = {{NIST Internet of Things (IoT) Program}},
  title        = {{NISTIR 8259: Foundational Cybersecurity Activities for IoT Device Manufacturers}},
  year         = {2020},
  howpublished = {\url{https://www.nist.gov/publications/foundational-cybersecurity-activities-iot-device-manufacturers}},
  note         = {National Institute of Standards and Technology,Accessed: January 21, 2026}
}

@misc{etsi303645,
  author       = {{ETSI (European Telecommunications Standards Institute)}},
  title        = {{ETSI EN 303 645: Cyber Security for Consumer Internet of Things}},
  year         = {2020},
  howpublished = {\url{https://www.etsi.org/standards/303645}},
  note         = {European Telecommunications Standards Institute,Accessed: January 21, 2026}
}

@misc{WikiDevi,
  title={WikiDevi},
  author={WikiDevi Contributors},
  year={2024},
  url={https://wikidevi.wi-cat.ru/},
  howpublished ={\url{https://wikidevi.wi-cat.ru/}},
  note={Accessed: January 21, 2026}
}

@misc{TechInfoDepot,
  title={TechInfoDepot},
  author={TechInfoDepot Contributors},
  year={2024},
  url={https://techinfodepot.shoutwiki.com/},
  howpublished={\url{https://techinfodepot.shoutwiki.com/}},
  note={Accessed: January 21, 2026}
}

@misc{OpenWrt,
  title={OpenWrt Project},
  author={OpenWrt Contributors},
  year={2024},
  url={https://openwrt.org/},
  howpublished = {\url{https://openwrt.org/}},
  note={Accessed: January 21, 2026}
}

@misc{codelinaro2026qsdk,
  title        = {QCA Software Development Kit (QSDK) Overview},
  author       = {{CodeLinaro Wiki}},
  howpublished = {\url{https://wiki.codelinaro.org/en/clo/qsdk/overview}},
  year         = {2026},
  note         = {Accessed: 2026-02-02},
  url          = {https://wiki.codelinaro.org/en/clo/qsdk/overview},
}

@misc{eu_cra,
  author       = {{European Union}},
  title        = {{Cyber Resilience Act (CRA)}},
  year         = {2022},
  note         = {Accessed: 2026-02-02},
  howpublished = {\url{https://digital-strategy.ec.europa.eu/en/policies/cyber-resilience-act}},
  note         = {European Union Regulation}
}

@online{fortinet_shadowv2_2025,
  title   = {ShadowV2 casts a shadow over IoT devices},
  author  = {Li, Vincent},
  year    = {2025},
  month   = nov,
  day     = {26},
  url     = {https://www.fortinet.com/blog/threat-research/shadowv2-casts-a-shadow-over-iot-devices},
  howpublished={\url{https://www.fortinet.com/blog/threat-research/shadowv2-casts-a-shadow-over-iot-devices}},
  note    = {Fortinet Blog (Threat Research). Accessed: 2026-01-23}
}

@online{securityscorecard_wrthug_2025,
  title   = {Operation WrtHug, The Global Espionage Campaign Hiding in Your Home Router},
  author  = {Friedenreich Maizles, Gilad and Kareem, Marty},
  year    = {2025},
  month   = nov,
  day     = {19},
  howpublished = {\url{https://securityscorecard.com/blog/operation-wrthug-the-global-espionage-campaign-hiding-in-your-home-router/}},
  note    = {SecurityScorecard Blog / STRIKE. Accessed: 2026-01-23}
}

@online{tp_link_quad7_2025,
  title   = {Technical News and Reports about Quad 7 (7777) Botnet aka CovertNetwork-1658},
  author  = {{TP-Link}},
  year    = {2025},
  month   = aug,
  day     = {29},
  howpublshed = {\url{https://www.tp-link.com/us/support/faq/4365/}},
  note    = {TP-Link Support FAQ (Last Updated: 2025-08-29). Accessed: 2026-01-23}
}

@online{bleepingcomp_raptor_train_2024,
  title   = {Chinese botnet infects 260,000 SOHO routers, IP cameras with malware},
  author  = {Ilascu, Ionut},
  year    = {2024},
  month   = sep,
  day     = {18},
  note    = {BleepingComputer. Accessed: 2026-01-23}, 
  howpublished = {\url{https://www.bleepingcomputer.com/news/security/flax-typhoon-hackers-infect-260-000-routers-ip-cameras-with-botnet-malware/}}
}

@misc{thez-labs_linux-exploit-suggester,
  author       = {The-Z-Labs},
  title        = {Linux Exploit Suggester},
  year         = {2023},
  url          = {https://github.com/The-Z-Labs/linux-exploit-suggester},
  howpublished={\url{https://github.com/The-Z-Labs/linux-exploit-suggester}},
  note         = {Accessed: 2026-02-02}
}

@misc{cvehound-Github,
  author       = {Evdenis},
  title        = {Cvehound},
  year         = {2023},
  url          = {https://github.com/evdenis/cvehound},
  howpublished={\url{https://github.com/evdenis/cvehound}},
  note         = {Accessed: 2026-02-02},
}

@misc{coccinelle_website,
  author       = {Coccinelle Team},
  title        = {Coccinelle},
  year         = {2023},
  url          = {https://coccinelle.gitlabpages.inria.fr/website/},
  howpublished={\url{https://coccinelle.gitlabpages.inria.fr/website/}},
  note         = {Accessed: 2026-02-02},
}

@misc{nvd:CVE-2024-1151,
  author       = {{National Vulnerability Database (NIST)}},
  title        = {{NVD - CVE-2024-1151}},
  howpublished = {\url{https://nvd.nist.gov/vuln/detail/CVE-2024-1151}},
  note         = {Accessed: 2026-02-02}
}

@misc{nvd:CVE-2024-22386,
  author       = {{National Vulnerability Database (NIST)}},
  title        = {{NVD - CVE-2024-22386}},
  howpublished = {\url{https://nvd.nist.gov/vuln/detail/CVE-2024-22386}},
  note         = {Accessed: 2026-02-02}
}

@misc{nvd:CVE-2023-6932,
  author       = {{National Vulnerability Database (NIST)}},
  title        = {{NVD - CVE-2023-6932}},
  howpublished = {\url{https://nvd.nist.gov/vuln/detail/CVE-2023-6932}},
  note         = {Accessed: 2026-02-02}
}

@misc{openwrt-github,
  author       = {{OpenWrt Project}},
  title        = {OpenWrt: Open Source Router and Embedded Operating System},
  howpublished = {\url{https://github.com/openwrt/openwrt/}},
  year         = {2026},
  note         = {Accessed: 2026-02-03}
}

@misc{foxconn-website,
  author       = {{Hon Hai Technology Group (Foxconn)}},
  title        = {Home - Hon Hai Technology Group - Foxconn},
  howpublished = {\url{https://www.foxconn.com/en-us}},
  note         = {Accessed: 2026-02-05}
}

@misc{mediatek-website,
  author       = {{MediaTek Inc.}},
  title        = {MediaTek | Home Page},
  howpublished = {\url{https://www.mediatek.com/}},
  note         = {Accessed: 2026-02-05}
}

@misc{qualcomm-website,
  author       = {{Qualcomm Incorporated}},
  title        = {Qualcomm: Intelligent Computing Everywhere},
  howpublished = {\url{https://www.qualcomm.com/}},
  note         = {Accessed: 2026-02-05}
}

@misc{freshtomato,
  title        = {{FreshTomato}: Enhanced Firmware for Broadcom-based Routers},
  howpublished = {\url{https://www.freshtomato.org/}},
  note         = {Accessed: Jan. 26, 2026},
  year         = {2026}
}

@misc{ddwrt,
  title        = {{DD-WRT}: Linux-based Open Source Router Firmware},
  howpublished = {\url{https://dd-wrt.com/}},
  note         = {Accessed: Jan. 26, 2026},
  year         = {2026}
}

@misc{securityweek-dlink-discontinued-2026,
  author       = {Arghire, Ionut},
  title        = {{Hackers Exploit Zero-Day in Discontinued D-Link Devices}},
  year         = {2026},
  month        = {1},
  howpublished = {\url{https://www.securityweek.com/hackers-exploit-zero-day-in-discontinued-d-link-devices/}},
  note         = {SecurityWeek, Jan. 7, 2026. Accessed: Feb. 6, 2026}
}

@misc{fbi-eol-proxies-2025,
  author       = {Federal Bureau of Investigation (FBI)},
  title        = {{Cybercriminal Proxy Services Exploiting End-of-Life Routers}},
  year         = {2025},
  month        = {5},
  howpublished = {\url{https://www.fbi.gov/investigate/cyber/alerts/2025/cybercriminal-proxy-services-exploiting-end-of-life-routers}},
  note         = {Public Service Announcement, May 7, 2025. Accessed: Feb. 6, 2026}
}

@misc{cisa_aa22_054a,
  author       = {{Cybersecurity and Infrastructure Security Agency (CISA)}},
  title        = {{AA22-054A}: New Sandworm Malware Cyclops Blink Replaces VPNFilter},
  howpublished = {\url{https://www.cisa.gov/news-events/cybersecurity-advisories/aa22-054a}},
  year         = {2022},
  note         = {Accessed: Jan. 26, 2026}
}

@INPROCEEDINGS{11023504,
  author={Ali, Mutahar and Arunasalam, Arjun and Farrukh, Habiba},
  booktitle={2025 IEEE Symposium on Security and Privacy (SP)}, 
  title={Understanding Users' Security and Privacy Concerns and Attitudes Towards Conversational AI Platforms}, 
  year={2025},
  volume={},
  number={},
  pages={298-316},
  doi={10.1109/SP61157.2025.00241}}

@techreport{ftc_mobile_security_updates,
  author      = {{Federal Trade Commission}},
  title       = {Mobile Security Updates: Understanding the Issues},
  institution = {U.S. Federal Trade Commission},
  year        = {2018},
  month       = feb,
  url         = {https://www.ftc.gov/reports/mobile-security-updates-understanding-issues}
}

@misc{ftc_blu,
  author       = {{Federal Trade Commission}},
  title        = {Mobile Phone Maker {BLU} Reaches Settlement with {FTC} over Deceptive Privacy and Data Security Claims},
  howpublished = {FTC Press Release},
  year         = {2018},
  month        = apr,
  url          = {https://www.ftc.gov/news-events/news/press-releases/2018/04/mobile-phone-maker-blu-reaches-settlement-ftc-over-deceptive-privacy-data-security-claims}
}

@misc{us_cyber_trust_mark,
  author       = {{Federal Communications Commission}},
  title        = {Cybersecurity Labeling for Internet of Things},
  howpublished = {Report and Order and Further Notice of Proposed Rulemaking, FCC 24-26},
  year         = {2024},
  month        = mar,
  url          = {https://www.fcc.gov/CyberTrustMark}
}

@misc{uk_csr_bill,
  author       = {{UK Parliament}},
  title        = {Cyber Security and Resilience (Network and Information Systems) Bill},
  howpublished = {House of Commons Bill},
  year         = {2026},
  url          = {https://bills.parliament.uk/bills/4035}
}

@inproceedings{utz2019uninformed,
  author    = {Utz, Christine and Degeling, Martin and Fahl, Sascha and Schaub, Florian and Holz, Thorsten},
  title     = {(Un)informed Consent: Studying {GDPR} Consent Notices in the Field},
  booktitle = {Proceedings of the 2019 ACM SIGSAC Conference on Computer and Communications Security},
  year      = {2019},
  pages     = {973--990},
  publisher = {ACM},
  doi       = {10.1145/3319535.3354212}
}
